\documentclass[a4paper,twocolumn,11pt, accepted=2024-08-15]{quantumarticle}
\pdfoutput=1
\usepackage[utf8]{inputenc}
\usepackage[english]{babel}
\usepackage[T1]{fontenc}
\usepackage{amsmath}
\usepackage{amssymb}
\usepackage{amsfonts}
\usepackage{amsthm}
\usepackage{bm}
\usepackage{dsfont}
\usepackage{hyperref}
\usepackage{amsmath}
\usepackage{amsthm}
\usepackage{amssymb}
\usepackage{graphicx}
\usepackage{esint}
\usepackage{titlesec}
\usepackage{amsfonts}
\usepackage{xcolor}
\usepackage[utf8]{inputenc}
\usepackage{amsfonts}
\usepackage{amsmath}
\usepackage{color}
\usepackage{footmisc}
\usepackage{bbold}
\usepackage{amsthm}
\usepackage{graphicx}
\usepackage{halloweenmath} 
\usepackage{curve2e} 
\usepackage{titlesec}
\usepackage{hyperref}

\usepackage{tikz}
\usepackage{lipsum}
\usepackage{xspace}

\usepackage[numbers]{natbib}

\newcommand{\parti}[0]{\mathbf{\mathcal{K}}}
\newcommand{\setparti}[0]{\mathcal{K}}  
\newcommand{\sizeparti}[0]{K}  

\newcommand{\vect}[1]{\boldsymbol{#1}}
\newcommand{\ket}[1]{\left| {#1}\right\rangle}
\newcommand{\bra}[1]{\left\langle {#1}\right|}
\newcommand{\braket}[2]{\left\langle {#1}\middle|{#2} \right\rangle}
\newcommand{\braAket}[3]{\left\langle {#1}\middle|{#2}\middle|{#3} \right\rangle}
\newcommand{\op}[1]{\hat{#1}}
\newcommand{\perm}[1]{\text{perm}\left({#1}\right)}

\newcommand{\Smatrix}{S}
\newcommand{\tvd}[1]{\text{tvd}({#1})}

\newcommand{\valery}{Shchesnovich\xspace}
\DeclareMathOperator*{\E}{\mathbb{E}}
\newtheorem{thm}{Theorem}

\begin{document}

\title{Efficient validation of Boson Sampling from binned photon-number distributions}

\author{Benoit Seron}
\email{benoitseron@gmail.com}
\affiliation{Quantum Information and Communication, Ecole polytechnique de Bruxelles, CP 165/59, Universit\'e libre de Bruxelles (ULB), 1050 Brussels, Belgium}

\author{Leonardo Novo}
\email{leonardo.novo@inl.int}
\affiliation{Quantum Information and Communication, Ecole polytechnique de Bruxelles, CP 165/59, Universit\'e libre de Bruxelles (ULB), 1050 Brussels, Belgium}
\affiliation{International Iberian Nanotechnology Laboratory (INL), Av. Mestre José Veiga, 4715-330 Braga, Portugal}

\author{Alex Arkhipov}
\email{arkhipov@alum.mit.edu}
\affiliation{Independent researcher}

\author{Nicolas J. Cerf}
\email{ncerf@ulb.ac.be}
\affiliation{Quantum Information and Communication, Ecole polytechnique de Bruxelles, CP 165/59, Universit\'e libre de Bruxelles (ULB), 1050 Brussels, Belgium}

\begin{abstract}
In order to substantiate claims of quantum computational advantage, it is crucial to develop efficient methods for validating the experimental data. We propose a test of the correct functioning of a boson sampler with single-photon inputs that is based on how photons distribute among partitions of the output modes. Our method is versatile and encompasses previous validation tests based on bunching phenomena, marginal distributions, and even some suppression laws. We show via theoretical arguments and numerical simulations that binned-mode photon number distributions can be used in practical scenarios to efficiently distinguish ideal boson samplers from those affected by realistic imperfections, especially partial distinguishability of the photons. 
\end{abstract}


\maketitle
 
\section{Introduction}

An important milestone in the field of quantum computing is the construction of a quantum device that can surpass even the most advanced classical super-computers at a specific task \cite{preskill2018quantum, preskill2021quantum}. For the purposes of demonstrating quantum computational advantage with near-term quantum devices, one of the problems that has been intensely investigated is that of Boson sampling. In their seminal paper \cite{bosonsampling}, Aaronson and Arkhipov presented strong complexity theoretic arguments showing that the task of sampling the output of a linear interferometry process involving many single photons is likely to be intractable for classical computers. Boson sampling sparked a great interest over the last decade and various alternative schemes were constructed, such as Scattershot boson sampling, Gaussian Boson sampling and others, in order to facilitate experimental implementations \cite{lund2014scattershot, Hamilton_2017, chakhmakhchyan2017boson, chabaud2017continuous}. 
These efforts culminated in multiple claims of quantum computational advantage with Gaussian Boson sampling \cite{zhong2020science, zhong2021phase, madsen2022quantum}, while standard boson sampling saw experimental implementations with $n = 20$ photons in $m = 60$ modes \cite{wang2019boson}. Other experimental platforms than photonics were also considered \cite{robens2022boson}.

Crucially, experimental implementations are unavoidably subject to different noise sources, such as those induced by partial distinguishability or particle loss, which may compromise claims of quantum computational advantage. Indeed, if the amount of noise is too large, then classical algorithms can sample from the outcome distribution efficiently \cite{rahimi2016sufficient, RenemaOxfordPaper, oszmaniec2018classical, garcia2019simulating, brod2020classical,renema2018classical, shchesnovich2019noise}.
Therefore, a thin line exists between the regime of classical computational hardness and efficient classical simulability. 

It is therefore of highest importance to develop efficient methods to discriminate an ideal boson sampler from a noisy one. However, the very formulation of the task at hand, which involves sampling from an exponentially large set of possibilities, makes the problem of verifying that the device is working properly highly non-trivial.  Ideally, one would like to assert that the experiment is generating samples from a distribution that is close enough to the ideal one simply by post-processing the classical data generated by the experiment. However, due to the flatness of the boson sampling distribution, this requires exponentially many samples \cite{hangleiter_sample_complexity}. Efficient verification schemes that guarantee closeness to the ideal distribution exist, but they require an active control over the experiment via the ability to do Gaussian measurements on the ouput states ~\cite{chabaud2021efficient}.   
Upon reasonable physical assumptions about the nature of noise \cite{goldstein2017decoherence, rohde2012error, kalai2014Gaussian,leverrier2013analysis, shchesnovich2014sufficient, arkhipov2015bosonsampling, aaronson2016bosonsampling, latmiral2016towards}, and forgetting about adversarial scenarios, we can restrict ourselves to the easier task of validation. It consists of verifying that the experiment passes some easy-to-check tests, which a boson sampler working in the quantum supremacy regime is expected to pass. These tests should be sufficiently sensitive to noise, so we can efficiently discriminate an ideal boson sampler from a noisy one. This is the main question we consider in this paper.

In our work we propose a simple validation scheme of boson sampling  based on jointly counting the number of photons in binned-together output modes. We give numerical evidence that these binned distributions are sensitive to typical sources of noise such as photon distinguishability, even for a small number of bins. At the same time, for a fixed choice of binning with a constant number of bins, we show that there is an efficient classical algorithm that approximates the corresponding binned distribution coming from an ideal boson sampler (Theorem~\ref{thm:approximate_TVD} in Sec.~\ref{sec:formalism}).  This way, it is possible to efficiently compare the data coming from an experiment or a mock-up sampler to that of an ideal sampler. This provides an advantage over other validation tests such as, for example, Bayesian approaches ~\cite{bentivegna2015bayesian,dai2020bayesian,flamini2020validating}, which require exponentially hard classical computations. Additionally, while other commonly used validation tests based on correlators or marginal distributions, which require only polynomial-time classical computations, can be spoofed by classical algorithms \cite{villalonga2021efficient}, we argue that spoofing binned distribution tests is much harder, since there are exponentially many choices of bins and sufficient variability between the distributions corresponding to different possible choices. We leave as an open question whether a clever efficient classical algorithm exists that can spoof the validation method proposed in our work.  

In what follows we introduce a list of desiderata for scalable validation tests for boson sampling, explaining how the validation scheme based on binned output modes fulfills them and detailing how the method stands in comparison with other methods in the literature.

\subsection{Validation of Boson Sampling}\label{sec:validation_criteria}
A plethora of validation tests for boson samplers have been proposed which are able to discriminate between ideal boson samplers and other mock-up distributions, such as the uniform distribution, distributions generated by distinguishable input photon,  or mean-field samplers \cite{walschaers2020signatures}. Techniques such as pattern recognition or machine learning~\cite{agresti2019pattern, flamini2019visual}, Bayesian testing \cite{bentivegna2015bayesian}, coarse-grained measurements \cite{wang2016certification, carolan2015universal}, Heavy Output Generation \cite{zhong2020science},  or the analysis of marginal distributions have also been applied \cite{villalonga2021efficient}. Each method offers its own advantages and disadvantages.
A recent publication combined a variety of the tests cited the above applied sequentially to come up with a single metric describing the quality of the experiment, the Photonic Quality Factor \cite{mezher2022assessing}.
In the context of our work, we put forward the following list of desiderata for a faithful validation test, focusing on sensitivity to realistic noise sources and computational efficiency. We would like a validation test to obey the following criteria: 
\begin{enumerate}
    \item \emph{Generality:} The interferometer in a boson sampling experiment is drawn at random from the Haar measure, so an important requirement for a validation test is that it is applicable to an \emph{arbitrary} linear interferometer.  
    \item \emph{Sensitivity to multiphoton interference:}  High-order multiphoton interference is at the core of the classical hardness of boson sampling \cite{shchesnovich2019noise, shchesnovich2021distinguishing}. Partial distinguishability of the input photons is one of the most important noise sources in boson sampling, which may render the outcome probabilities easy to approximate. Experiments with a constant amount of photon distinguishability may be simulated by considering only $k$-photon interference terms, for some fixed value of $k$ \cite{RenemaOxfordPaper, moylett2019classically}. Hence, a validation test should be sensitive to high-order multiphoton interference in order to discriminate an ideal boson sampler from one with partially distinguishable photons.
    \item \emph{Sampling efficiency:} Resources available for validation are limited since, compared to the exponentially large system size (domain of the sampled distribution), only a moderate amount of samples are observed. For this reason, we request that a validation test is able to discriminate a noisy boson sampler (with some fixed noise parameters) from an ideal one by using a number of samples that scales only polynomially in the system size.  
    \item \emph{Computational efficiency:} A last requirement is efficiency in post-processing the classical data. Many validation tests that aim at comparing the experiment to an ideal boson sampler require the computation of outcome probabilities, or the classical simulation of ideal boson sampling, which takes exponential time. This becomes prohibitive as experiments grow larger, which is why we restrict ourselves to polynomial-time computations in the system size.   
\end{enumerate}

Additionally, it is important to remark that, in a realistic setting, photon loss also plays a prominent role. Unlike partial distinguishability, the amount of loss is easy to estimate from data coming from the experiment (or even tests with classical light). We can assume that a lossy boson sampling experiment that aims at demonstrating quantum computational advantage is able to obtain high-enough output photon counts such that, if the rest of the experiment was ideal, it would still surpass the best classical simulation algorithms. As most of the outcomes would correspond to events with lost photons, it is important that a validation test is able to use this data in order to diagnose \emph{other sources of noise}, such as partial distinguishability, which may render the experiment classically simulable. If validation required considering only postselected outcomes with no lost photons, this would sharply decrease the sampling rate of usable events, possibly making the validation unfeasible in a reasonable amount of time. We shall come back to this point later in this work.

\subsection{Our contribution}

\begin{figure*}[t]
\onecolumn
  \centering
  \includegraphics[width=\textwidth]{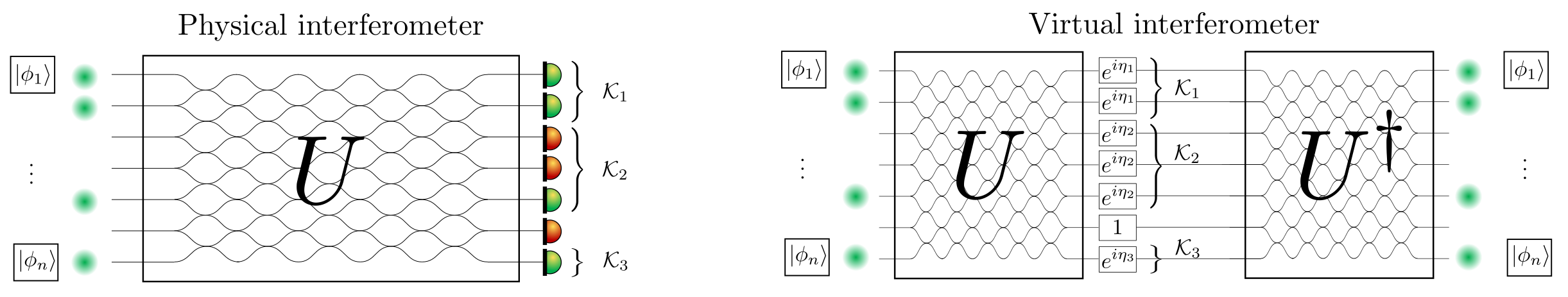}
  \caption{(Left) \textbf{Interferometric setup.} Single photons are sent through the first $n$ input modes of a $m$-mode linear interferometer $U$. Each photon at input $j$ carries internal degrees of freedom (polarization, arrival time, etc.) described by an (internal) wavefunction $\ket{\phi_j}$. To validate the device we use the probability distribution obtained by counting photons in binned-together output modes. In the figure, we represent in green the detectors that have detected one photon and in red the detectors that did not click. In this example, we have 3 bins and the outcome observed is $\vec{k}=(2,1,1)$ (the first subset $\mathcal{K}_1$ observes two photons, $\mathcal{K}_2$ observes a single photon, etc.). Such outcome probabilities are sensitive to partial distinguishability of the input and can be used to discriminate an ideal boson sampler from a noisy one. (Right) \textbf{Virtual interferometer.} The binned output mode probabilities can be obtained from the 
  characteristic function of this distribution. The latter can be interpreted as a probability amplitude in a virtual interferometry process. More precisely, computing $x(\vect{\eta})$ as described in Eq. \eqref{eq:xl_perm} (through Eq. \eqref{eq:characteristic_interferometer}) is equivalent to computing the amplitude of the process where the input state of the boson sampler is left unchanged while going through a virtual interferometer $V$ built by sandwiching the physical interferometer $U$ and its hermitian conjugate $U^\dagger$ with a diagonal matrix of phases, encoding the choice of partition (as defined in Eq. \eqref{eq:defV} to \eqref{eq:dab}).}
  \label{fig:scheme_verification}
\end{figure*}
\twocolumn

We propose a validation scheme for Boson Sampling which aims at fulfilling the list of requirements stated above. Our scheme (Sec.~\ref{sec:formalism}) is based on a simple coarse-graining of the data coming from the boson sampling device: we group the output modes into different subsets and count how many photons end up in each of them. Our outcomes are thus given by a vector $\vect{k} = (k_1,\dots, k_{\sizeparti})$, where $k_z$ is the number of photons observed in subset $\setparti_z$ (see left panel of Fig.~\ref{fig:scheme_verification}). This \emph{binning} of the output modes into different subsets allows us to deal with a space of events of much smaller size than the exponentially many outcomes of the boson sampler. For a fixed number $\sizeparti$ of subsets, the number of possible configurations of the photons in the bins is bounded by $(n+1)^K$, where $n$ is the photon number. This implies that some of the probabilities are relatively large (of size $1/\text{poly}(n)$), and thus a meaningful estimation (i.e.~up to relative error) of these large probabilities can be obtained from a polynomial number of experimental runs. Crucially, we show that a classical algorithm exists that can also estimate these probabilities efficiently, not only for ideal boson samplers but also noisy ones, involving partially distinguishable input photons as well as loss.

The validation test we consider is based on comparing these theoretically predicted probabilities to the experimentally observed ones. We provide analytical and numerical evidence that it is possible to use binned output distributions to efficiently discriminate between bosonic and classical input particles, as well as some models of partial distinguishability. We also argue that this way of validating boson samplers is sensitive to partial distinguishability, even in the presence of small amounts of loss.

\subsection{Comparison to other existing methods}

The versatility of the method we consider lies in the fact that it can be used for any interferometer and that the choice of the bins is completely arbitrary. It can even be done after the experiment -- the same data can be tested using multiple choices of subsets, possibly ones chosen randomly. This versatility allows us to connect this method to some important validation tests for standard boson samplers that have been suggested in previous literature and even retrieve some of them as particular cases. 

\subsubsection{Correlators and marginal distributions}

One of the most common validation protocol for boson samplers relies on low-order correlation functions of the output mode counts $\op{n}_i$ \cite{walschaers2016statistical, walschaers2018statistical} such as
\begin{align}
    c_i &= \langle n_i\rangle \label{eq:first_order_correlator} \\
    c_{ij} &= \langle n_i n_j\rangle - \langle n_i\rangle \langle n_j\rangle \label{eq:second_order_correlator}
\end{align}
Correlations of order $3$ and $4$ have also been considered \cite{giordani2018experimental, zhong2021phase}. 

A closely related validation method is that of computing $k$-marginals of the boson sampling distributions of an ideal experiment -- which results from looking at the photon counts coming from a constant number $k$ of detectors -- and comparing them to the experiment itself \cite{villalonga2021efficient}. This data can be used to estimate the correlators mentioned above. While marginals of fixed size and low-order correlators can be computed in polynomial time, they are known to be insensitive to higher order multiphoton interferences. As previously mentioned, the latter are crucial to reproduce the boson sampling distribution to sufficient precision~ \cite{shchesnovich2022boson}. In fact, it is possible to construct efficient classical mock-up samplers that are consistent with all marginals of order $k$ \cite{villalonga2021efficient}. Therefore, this scheme alone cannot be used to justify claims of quantum computational advantage. 

We note that marginals distribution can be recovered as a particular case of our scheme, as it corresponds to choosing $\sizeparti = k$ subsets with a single output mode in each. Our scheme however can be adapted to be sensitive to higher order interferences: if, for example, two equal-sized subsets are chosen, the way photons distribute in this partition of the output modes cannot be well approximated by taking into account only few-photon interference terms.

\subsubsection{Full bunching}

\valery presents an interesting scheme that aims at fulfilling all the requirements stated in Sec.~\ref{sec:validation_criteria} \cite{shchesnovich2016universality, shchesnovich2021distinguishing}. 
It relies on observing full bunching in a subset of the output modes, i.e.~all the input photons are found in some chosen subset. Equivalently,  one can focus on observing no photon in the complementary subset \cite{shchesnovich2021distinguishing}.

Full bunching probabilities can be approximated efficiently and numerical simulations predict that, for Haar random matrices, this quantity is maximized when photons are fully indistinguishable and decreases when they are partially distinguishable. While explicit counter-examples to this general rule of thumb were demonstrated in \cite{seronBosonBunching}, going against general physical intuition, the method remains practical as it holds very well on average, independently of which subset is chosen. 

While there is evidence that validation using bunching probabilities ticks all the boxes of desirable properties put forward in Sec.~\ref{sec:validation_criteria}, we improve on this method quantitatively. In fact, the full bunching probability in a subset can be seen as a particular outcome probability of our more general scheme, which takes into account full photon counting distribution in one or more subsets. This allows for better distinguishing power, requiring fewer experimental samples for the validation task.

\subsubsection{Suppression laws}

Certain validation tests rely on a specific choice of optical network \cite{tichy2014stringent}. Symmetries in the unitary matrix lead to suppression laws where many outputs are prohibited \cite{dittel2018totally, viggianiello2018experimental, crespi2015suppression}. These methods are interesting tools to diagnose noise in the network or input state, but are restricted to only a handful of networks, such as the Fourier or Sylvester matrices. However, the arguments for the hardness of boson sampling require the unitary to be chosen randomly according the the Haar measure \cite{bosonsampling}. If one can precisely tune a reconfigurable network \cite{madsen2022quantum}, then validation can be executed first through suppression laws, and then the network can be changed to a random unitary to obtain samples. The main drawbacks are that this method leaves the door open for errors to appear in the samples when the circuit is reconfigured and provides no way to validate the final samples after circuit reconfiguration. 

Nevertheless, interferometers possessing some symmetries such as the Fourier interferometer are interesting devices to test multiphoton interference due these suppression laws and their sensitivity to distinguishability of the photons. We show in Sec.~\ref{sec:Fourier} that our formalism allows us to compute analytically some binned output distributions for Fourier interferometers, revealing striking differences between the behavior of distinguishable particles and ideal bosons. We also observe that some characteristic suppressions observed for ideal bosons are inherited by the binned output distribution, for an appropriate choice of the bins.

\subsubsection{Validation from coarse-grained measurements}

Other proposals exist to validate boson samplers using coarse-grained data. In \cite{wang2016certification}, the authors classify observed events in bubbles in the state space (unlike our binning of output spatial modes). Bubbles are constructed iteratively and are centered around high probability events. Numerical evidence is given to show that this is a good validation method against, for example, distinguishable input photons. However, with this validation method, the comparison of the experiment to an ideal boson sampler would still require a classical simulator of the latter which would not be efficient. A similar statement also applies to pattern recognition techniques \cite{agresti2019pattern}.
\subsection{Structure of the paper}
The core of our work is divided into three sections followed by a discussion section. In Sec.~\ref{sec:formalism}, we explain in detail the mathematical techniques to compute the binned output distribution generated by ideal or noisy boson samplers. We focus on the analysis of the complexity of the method, showing that efficient approximations of the distribution can be obtained. In Sec.~\ref{sec:Fourier}, we present analytical results for binned distributions in Fourier interferometers. This section may be skipped entirely by a reader who is only interested in our results regarding the validation of experiments with Haar-random interferometers. The latter are presented in our main results section (Sec.~\ref{sec:validation}). We conclude with a discussion section containing open questions and perspectives for future work.  

\section{Formalism}\label{sec:formalism}
As previously mentioned, the boson sampling validation method we consider in this work is based on how photons distribute into subsets of output modes, which we will also refer to as bins. 

We consider the partition of $\mathcal{M}=\{1,2,\dots, m\}$ into 
$\sizeparti$ non-empty and mutually disjoint subsets $\setparti_z\subset \mathcal{M}$ with $z\in \{1,\dots,\sizeparti\}$. If the photon configuration at the output of a boson sampler is $\vect{s}= (s_1, s_2, \dots, s_m)$, the way photons distribute in this partition denoted as $\parti=\{\setparti_1,\dots,\setparti_\sizeparti\}$ is fully defined by a vector $\vect{k}$ of dimension $\sizeparti$, whose components are given by  %
\begin{equation}
 k_z= \sum_{j\in \setparti_z} s_j .
\end{equation}
We are interested in computing the probabilities $P(\vect{k})$ of observing the different possible photon number configurations in this partition. In this section, we show a classical algorithm that, for an experiment with $n$ input photons and constant number of subsets $K$, estimates the probabilities $P(\vect{k})$ up to total variation distance $\beta$ in polynomial time $O(n^{2K+2}\log(n) \beta^{-2})$, given some theoretical model for the experiment which may include partial distinguishability between the photons as well as losses.  Our derivation is based on the approximation of the characteristic function of the distribution $P(\vect{k})$  and is inspired by a result of Arkhipov for approximating linear statistics of ideal boson samplers \cite{alexnotes}.  The main idea we use is that this characteristic function can be interpreted as a probability amplitude of a virtual interferometric process, as depicted in Fig.~\ref{fig:scheme_verification}, and thus it can be approximated via Gurvits randomized algorithm for permanent approximation \cite{bosonsampling}.

\subsection{Photon-counting probabilities in partitions}\label{subsec:formalism}
For a given partition $\parti=\{\setparti_1,\dots,\setparti_\sizeparti\}$, the probability of observing a certain photon number configuration $\vect{k}$ can be obtained by summing the probabilities of all outcomes of the boson sampler that are consistent with it. However, this is impractical -- each outcome probability is hard to compute (a tensor permanent if photons are partially distinguishable \cite{tichy2015_partial_distinguishability}), and there can be an exponentially large number of events that are consistent with a given photon number configuration in the partition.

A better way to compute these probabilities $P(\vect{k})$ is via the characteristic function associated with this distribution, defined as 
\begin{align}
    \label{eq:characteristicFunctionDef}
    x({\vect{\eta}})&= 
    \E_{\vect{k}}\left[ \exp\left(i  \vect{\eta}\cdot \vect{k}\right)\right]\\
    & = \sum_{\vect{k}\in \Omega^{K}}P({\vect{k}}) \exp\left( i \vect{\eta}\cdot \vect{k}\right), 
\end{align}
with ${\vect{\eta}} \in \mathbb{R}^K$. Here, we have defined the set 
\begin{equation}
    \Omega^{\sizeparti}= \{(k_1, k_2, \dots, k_{\sizeparti})~|~ k_z\in \Omega, \forall z\in \{1,\dots,\sizeparti\}\},
\end{equation}
with $\Omega= \{0,1,\dots,n\}$. It can be seen that the probabilities $P(\vect{k})$ can be retrieved by evaluating $x({\vect{\eta}})$ at $(n+1)^{\sizeparti}$ points on a $\sizeparti$-dimensional grid, namely,
\begin{align}
\vect{\nu_{\vect{l}}} = \frac{2 \pi  \vect{l}}{n+1} ,
\quad \text{with~} l_z\in \Omega, \forall z\in \{1,\dots,\sizeparti\}
\end{align}
and taking the multidimensional Fourier transform, i.e.  
\begin{align}\label{eq:Pk_from_xl}
    P(\vect{k})= \frac{1}{(n+1)^K}\sum_{\vect{l}\in \Omega^{\sizeparti}} x(\vect{\nu_{\vect{l}}}) \exp\left(-i \vect{\nu_{\vect{l}}} \cdot \vect{k}\right). 
\end{align}
To evaluate the characteristic function we consider the usual boson sampling setting where $n$ photons are sent through a linear interferometer of $m$ modes, with one photon occupying each of the first $n$ input modes (a more general input, with more than one photon per mode, is considered in Appendix \ref{appendix:amplitudes_partially_distinguishable}). In order to model partial distinguishability between photons, we assume the internal degrees of freedom of the photon entering mode $j$, such as polarization or spectral distribution, are described by an internal state $\ket{\phi_j}$. The input state can then be written as 
\begin{equation}\label{eq:psi_in_pd}
\ket{\Psi}_{in} = \prod_{j=1}^n  \left(\op{a}_{j, \phi_j} ^\dagger \right) \ket{0}
\end{equation}
where $\ket{0}$ is the vacuum state and $\op{a}_{j, \phi_j} ^\dagger$ is the creation operator corresponding to a photon in mode $j$ and internal state $\ket{\phi_j}$.
We also define a basis $\{ \ket{\Phi_j}\}$ for the internal Hilbert space of the photons such that
\begin{equation}
    \sum _j \braket{\phi_k}{\Phi_j}\braket{\Phi_j}{\phi_k}=1, \quad \forall k.
\end{equation} 
Note that, even though the internal Hilbert space of the photons may be of infinite dimension, we only need at most $n$ basis elements to span the Hilbert space generated by the $n$ states $\ket{\phi_j}$. In addition, since the basis $\{\ket{\Phi_j}\}$ is orthonormal, the operators $\hat{a}_{i,\Phi_j}$ and $\hat{a}^{\dagger}_{i,\Phi_j}$ obey the usual commutation relations $[\hat{a}_{i,\Phi_j}, \hat{a}^{\dagger}_{k,\Phi_l}]=\delta_{ik}\delta_{jl}$. Therefore, we can define the number operator, which counts the number of photons in a spatial mode independently of their internal states, as 
\begin{equation}
\hat{n}_i= \sum_j \hat{a}^{\dagger}_{i, \Phi_j} \hat{a}_{i,\Phi_j}. 
\end{equation}

Following the formalism of \cite{tichy2015_partial_distinguishability, shchesnovich2015partial}, we use the standard assumption that the interferometer $\hat{U}$ acts only on the spatial modes, leaving the internal wavefunctions untouched. The relation between input and output modes is hence described by an $m\times m$ unitary matrix $U$ via the equation
\begin{eqnarray}
\op{a}_{j, \phi_j} ^\dagger \rightarrow \op{b}^\dagger_{k, \phi_j} \mathrm{~~~with~~~}\nonumber \\
\op{a}_{j, \phi_j}^\dagger = \op{U} \op{b}_{j, \phi_j} ^\dagger \op{U}^\dagger = \sum_{k = 1}^m U_{jk} \, \op{b}^\dagger_{k, \phi_j}.
\end{eqnarray}
The operator which counts the number of photons in a given subset of output spatial modes is then 
\begin{equation}
    \op{N}_{\setparti_z} = \sum _{j \in \setparti_z} \op{n}_j=\sum _{j \in \setparti_z}\sum_k \hat{b}^{\dagger}_{j, \Phi_k} \hat{b}_{j,\Phi_k} . 
\end{equation}
We demonstrate in Appendix \ref{appendix:amplitudes_partially_distinguishable} that the characteristic function $x(\vect{\eta})$ can be computed as the quantum expectation value
\begin{align}
    x({\vect{\eta}})&= \bra{\Psi_\mathrm{out}}e^{i \vect{\eta}\cdot \vect{\hat{N}_{\mathcal{K}}}} \ket{\Psi_\mathrm{out}}\\
    ~&=\bra{\Psi_\mathrm{in}} \hat{U}^\dagger e^{i \vect{\eta}\cdot \vect{\hat{N}_{\mathcal{K}}}} \hat{U}\ket{\Psi_\mathrm{in}}\label{eq:characteristic_interferometer}, 
\end{align}
where we have used the notation
\begin{equation}
    \vect{\eta}\cdot \vect{\hat{N}_{\mathcal{K}}}= \sum_{z=1}^{\sizeparti} \eta_z \op{N}_{\setparti_z}.
\end{equation}
At this point, it is useful to note that $\op{V}(\vect{\eta})=\op{U}^{\dagger} e^{i \vect{\eta}\cdot \vect{\hat{N}_{\mathcal{K}}}} 
\op{U}$  is a linear interferometer characterized by an $m\times m$ unitary matrix $V(\vect{\eta})$ (see right panel of Fig.~\ref{fig:scheme_verification}). This matrix is constructed as 
\begin{align}\label{eq:defV}
V(\vect{\eta})&=U^{\dagger} \Lambda(\vect{\eta})U, 
\end{align} 
where $\Lambda(\vect{\eta})$ is a diagonal matrix given by a product of diagonal matrices 
\begin{equation}
\Lambda(\vect{\eta})= \prod_{z=1}^K D^{(z)}(\eta_z), 
\end{equation}
such that 
\begin{equation}
\label{eq:dab}
    D^{(z)}_{ab}(\eta_z)= \begin{cases}
    e^{i \eta_z},~\text{ if}~a=b~\text{and}~a\in\setparti_z,\\
    1, ~\text{ if}~a=b~\text{and}~a\notin\setparti_z,\\
    0, ~\text{ if}~a\neq b.
    \end{cases}
\end{equation}

Using the results of Ref.~\cite{bosonsampling}, it can be shown that in the ideal boson sampling scenario where all photons are indistinguishable, the computation of the characteristic function is given by a matrix permanent
\begin{equation}
    x(\vect{\eta}) = \perm{V_n(\vect{\eta})},
\end{equation}
where $V_n$ corresponds to the $n\times n$ upper left submatrix of the  matrix $V$. In the more general case where the input photons can have different internal wavefunctions (see Eq.~\eqref{eq:psi_in_pd}), this expression is modified in a simple way. By defining the Gram matrix 
\begin{equation}
\label{eq:S_matrix}
S_{ij}=\braket{\phi_i}{\phi_j}
\end{equation}
of the overlaps of the internal states of the photons, the expression takes the form 
\begin{equation}
\label{eq:xl_perm}
x(\vect{\eta})=\perm{S\odot V_n(\vect{\eta})}, 
\end{equation}
where $\odot$ is the Hadamard (elementwise) product: $(A\odot B)_{ij} = A_{ij}B_{ij}$. An explicit derivation of the expression is done in Appendix~ \ref{appendix:amplitudes_partially_distinguishable}. 

\subsection{Loss and dark counts}

\begin{figure}
  \centering
  \includegraphics[width=0.5\textwidth]{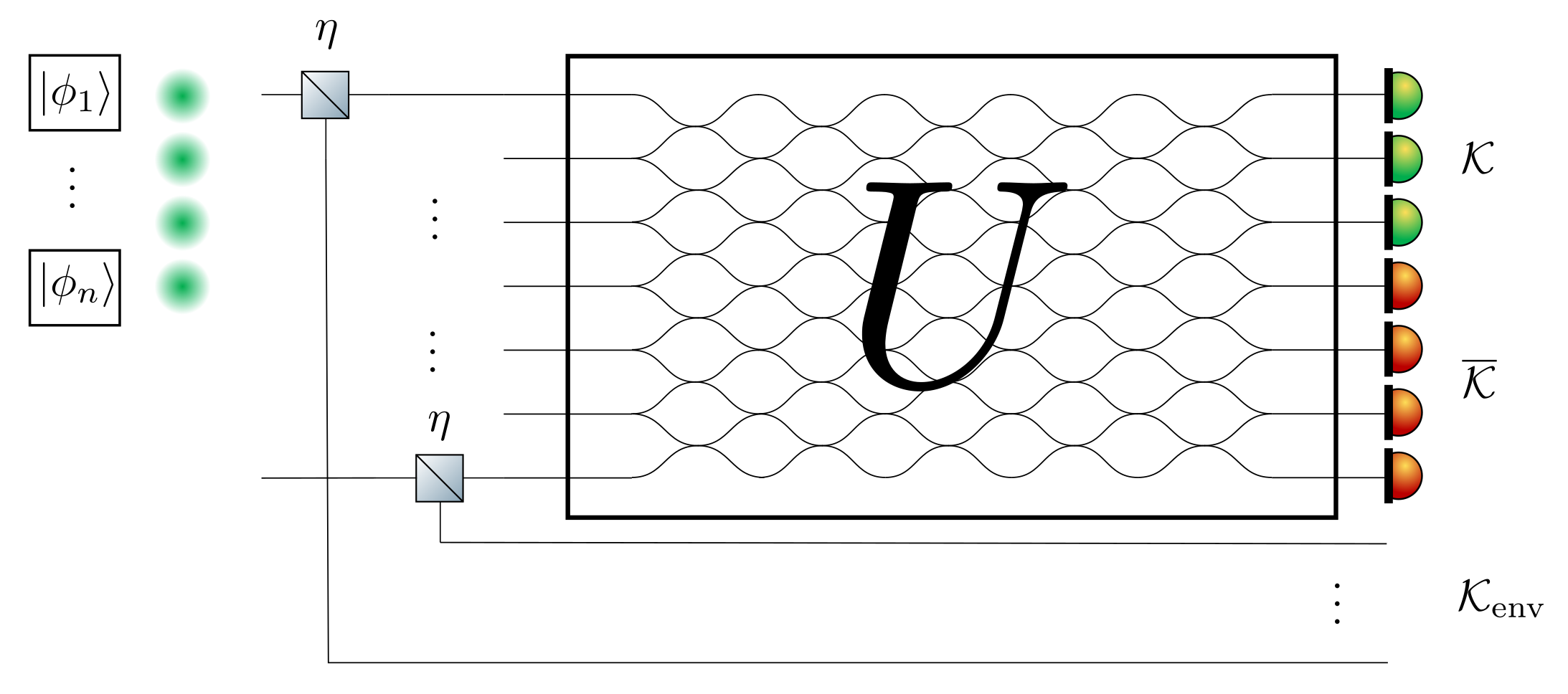}
  \caption{\textbf{Lossy interferometer model.} We use a simple model of uniform loss: each photon has a goes through a beam-splitter of transmissivity $\eta$ before entering the interferometer. If the photon is reflected, it is sent into an environment mode, which simulates a lost photon. The photon number distribution in binned modes of a lossy boson sampler can be computed by considering this larger lossless interferometer.}
  \label{fig:interferometer_lossy}
\end{figure}

We can accommodate photon loss at little extra cost with this formalism. In general, a lossy linear optical circuit can be described by first applying a lossless linear interferometer $W_1$, followed by $m$ parallel loss channels and a final lossless linear interferometer $W_2$ \cite{garcia2019simulating}. In turn, a loss channel acting on a given optical mode can be modelled in a \emph{unitary} way, by introducing an ancillary environment mode in the vacuum state and applying a beam-splitter with a given transmissivity $\lambda_i$. This implies that the output statistics of a lossy boson sampler of $m$ modes can be recovered by considering a larger lossless interferometer of $2m$ modes, described by a unitary matrix $\tilde{U}$, where only the first $m$ modes are measured. In the case of \emph{uniform loss}, the scheme can be simplified by considering an array of beam-splitters with the environment modes before the interferometer described by a unitary $U= W_1 W_2$ (see Fig.~\ref{fig:interferometer_lossy}).

Taking this into consideration, it is easy to adapt the formalism from Sec.~\ref{subsec:formalism} to obtain the photon-counting probabilities in a partition of the output modes of a lossy interferometer. We simply consider the distribution in a partition of the larger interferometer $\{\setparti_1,\dots,\setparti_K, \setparti_{env}\}$, where the last subset contains all the environment modes.  Note that the size of the matrix whose permanent we need to compute in Eq.~\eqref{eq:xl_perm} depends only on the number of input photons. Hence, even for lossy interferometers, the characteristic function of the photon number distribution in the binned output modes is still given by a permanent of an $n\times n$ matrix (see Eqs.~\eqref{eq:characteristic_interferometer} and \eqref{eq:xl_perm})), which in this case is a submatrix of a $2m \times 2m $ unitary matrix.

Another source of experimental noise is dark counts from the detectors. Even though we do not consider explicitly this effect in our work we note that the effect of dark counts can be incorporated in the calculation of the probabilities $P(\vect{k})$. Indeed, dark counts are a default of detectors, and do not change the underlying physics of the experiment. It suffices to add their statistical contribution on top of the of the experiment with no dark counts. For example, consider the simple case of a uniform dark count probability generation $p_d \ll 1$, and a single subset of size $K$, the probability of observing $k_d$ dark counts is given by a binomial 
\begin{equation}
    \label{eq:dark_count_binomial}
    q(k_d) = {K \choose k_d} p_d^{k_d}(1-p_d)^{K-k_d}
\end{equation}
The overall probability $P_d(k)$ of observing $k$ photons is thus the convolution of the original probability distribution $P(k)$ with \eqref{eq:dark_count_binomial}
\begin{equation}
    P_d(k) = (P \ast q)(k) = \sum _{k_d = 0}^k q(k_d) P(k-k_d)
\end{equation}
This has a moderate effect in the complexity of computing the probability distribution we are interested in. 

\subsection{Complexity analysis}
As shown in Eq.~\eqref{eq:Pk_from_xl}, the probabilities $P(\vect{k})$ can be evaluated by taking a multidimensional DFT of the values of the characteristic function $x(\vect{\eta})$. Using fast methods to compute the multidimensional DFT, the full distribution can be computed in time 
\begin{equation}\label{eq:complexity}
T = O(K (n+1)^{K} \log(n+1) C_x),    
\end{equation}
where $C_x$ is the cost of computing a single value of $x(\vect{\eta})$. Each quantity $x_{\vect{\eta}}$ requires the evaluation of a $n\times n$ permanent which can be computed exactly using Ryser's algorithm -- the best known classical algorithm for the exact computation of permanents -- in time $\mathcal{O}(n 2^n)$. However, in a practical scenario, the exact computation of the probabilities is not necessary since the experimental estimation of these probabilities will always carry an error due to the finite number of samples. Precisely, we need  $\mathcal{O}(1/\epsilon^2)$ samples to estimate the probabilities $P_{\text{exp.}}(\vect{k})$ up to an additive error $\epsilon$. Hence, if we assume we can run the experiment a polynomial number of times, we can only estimate the probabilities to a polynomially small error.  

In what follows, we show that classical algorithms 
can also efficiently obtain such polynomially small additive error approximations due to Gurvits' permanent approximation algorithm \cite{gurvits_2002_permanent_approx}. This algorithm allows for the approximation of permanents of unitary matrices up to error $\epsilon$ in time  $O(n^2/\epsilon^2)$. Our result regarding the computation of the approximate binned distribution up to a fixed \emph{total variation distance} $\beta$ is the following. 


\begin{thm}\label{thm:approximate_TVD}
For a constant partition size $\sizeparti$, there is a classical algorithm that computes an approximate distribution of probabilities $\tilde{P}(\vect{k})$ such that 
$$\sum_{\vect{k}}| \tilde{P}(\vect{k}) - P(\vect{k})|\leq \beta$$ in time $O(n^{2K+2}\log(n) \beta^{-2})$
\end{thm}
\begin{proof}
Consider an approximate distribution of probabilities $\tilde{P}(\vect{k})$ 
\begin{align}
\tilde{P}(\vect{k})  = \frac{1}{(n+1)^K}\sum_{\vect{l}\in \Omega^K} \tilde{x}(\nu_{\vect{l}})\exp\left(-i \nu_{\vect{l}}\cdot \vect{k}\right), \label{eq:approximateP}
\end{align}
obtained from the Fourier transform of approximate values of the characteristic function \begin{equation}
\tilde{x}(\nu_{\vect{l}})=x(\nu_{\vect{l}})+\epsilon_{\vect{l}},    
\end{equation} where $\epsilon_{\vect{l}}$ is an error term. It can be shown that 
\begin{align}
    \sum_{\vect{k}} |\tilde{P}(\vect{k})-P(\vect{k})|^2 = 
    & \sum_{\vect{k}} 
       \left|\sum_{\vect{l}} \frac{\epsilon_{\vect{l}}}{(n+1)^\sizeparti} e^{-i \nu_{\vect{l}}\cdot \vect{k}}\right|^2\\ \nonumber
       &= \frac{\sum_{\vect{l}} |\epsilon_{\vect{l}}|^2}{(n+1)^\sizeparti}, 
\end{align}
where we have used Parseval's theorem. By defining $\epsilon= \max_{\vect{l}} \epsilon_{\vect{l}}$, we can bound
the $\ell_2$-norm between the approximate and exact distribution by 
\begin{equation}
        \sqrt{\sum_{\vect{k}} |\tilde{P}(\vect{k})-P(\vect{k})|^2}\leq \epsilon. 
\end{equation}
This implies that the $\ell_1$-norm is bounded by 
\begin{equation}
        \sum_{\vect{k}} |\tilde{P}(\vect{k})-P(\vect{k})|\leq (n+1)^{\sizeparti/2} \epsilon. 
\end{equation}
Therefore, in order to obtain an $\ell_1$-norm bounded by some constant value $\beta$, we need to compute the approximate values $\tilde{x}(\nu_{\vect{l}})$ up to an error $\epsilon_{\vect{l}}\leq \epsilon\leq  \beta (n+1)^{-\sizeparti/2}$. The cost of evaluating each of these values using Gurvits algorithm is $C_x= O(n^{\sizeparti+2} \beta^{-2} )$. This can be seen as follows. For any distinguishability matrix $S$, we can evaluate $x(\vect{\eta})$ up to error $\epsilon||S\odot V_n(\vect{\eta})||^n$ in time $O(n^2/\epsilon^2)$ \cite{bosonsampling}. Due to a theorem by Schur (see section of Hadamard product from Ref.~\cite{johnson1990matrix}), we have that 
\begin{equation}
    ||S\odot V_n(\vect{\eta})||\leq (\max_i S_{ii})~ ||V_n(\vect{\eta})||\leq 1.
\end{equation}

 Finally, the complexity of obtaining the full approximate probability distribution $\tilde{P}(\vect{k})$ can be bounded using Eq.~\eqref{eq:complexity} by
\begin{equation}
    T= O(n^{2\sizeparti+2}\log(n) \beta^{-2}), 
\end{equation}
where we considered $\sizeparti$ to be a constant independent of $n$.
\end{proof}
This shows that the photon counting probabilities in the binned output modes can be approximated efficiently (in polynomial time) for any polynomially small additive error \footnote{We ignore the complexity of computing the matrix $V(\vect{\eta})$, which is polynomial in $m$, the number of modes, itself assumed to be polynomial in the number of photons $n$ in most use cases.}. While this fact is practically irrelevant to efficiently estimate usual boson sampling event probabilities (as they, on average, decrease exponentially, hence requiring $\epsilon$ to be exponentially small), it is relevant here as the probabilities $P(\vect{k})$ sum to one and that there are only $O((n+1)^K)$ of them. 

\subsubsection{Complexity of computing marginals}
The formalism we presented also allows us to compute marginal boson sampling distributions, a commonly used validation test for boson sampling experiments \cite{villalonga2021efficient}. For example, if we are interested in the distribution over the first $\sizeparti$ modes,   we take each subset to be a single mode, i.e.~$\setparti_z= \{z\}$, for $z\in \{1,\dots, \sizeparti\}$. It is known that marginal distributions of ideal boson sampler (with fully indistinguishable photons) can be computed \emph{exactly} in polynomial time \cite{bosonsampling, gurvits_marginals}. Here we show that the formalism we consider allows us to recover this result and extend it to any input (pure) state of partially distinguishable photons, as long as the internal states of the photons belong to a Hilbert space of constant size. The latter is given by the rank of the $S$ matrix (see Eq.~\eqref{eq:S_matrix}), which we denote as $r_S$. 

To show this result, we follow very similar lines to Ref.~\cite{gurvits_marginals} and use on the existence of an efficient algorithm to exactly compute the permanent of $n$-dimensional square matrices of the form $\mathds{1} + A$, where $A$ has some constant rank $r_A$. Precisely, this takes time $O(n^{2 r_A+1})$. In order to compute the marginal distribution we now consider the characteristic function (also called generating function) given by 
\begin{align}
x(\vect{\eta})&= \bra{\Psi_{in}} \hat{U}^\dagger e^{i \sum_{j=1}^r \eta_j \hat{n}_j} \hat{U}\ket{\Psi_{in}}\\&= \perm{V_n(\vect{\eta})\odot S}.    
\end{align}
In this case, we can write 
\begin{equation}
    V_n(\vect{\eta})\odot S= \mathds{1}_n + W(\vect{\eta})\odot S,
\end{equation}
where $\mathds{1}_n$ is the identity matrix of dimension $n$ and 
\begin{align}
    W(\vect{\eta})=  U^\dagger (\Lambda(\vect{\eta})-\mathds{1}_n ) U.   
\end{align}
It is possible to see that $W(\vect{\eta})$ is a matrix of rank $r$ and thus $\text{rank}(W(\vect{\eta})\odot S)\leq K  r_S$.   This way, we can bound the cost of exactly calculating the generating function of the marginal distribution corresponding to a subsystem of $K$ output modes by 
\begin{equation}
    C_x= O(n^{2 K r_S +1}).
\end{equation}
This result can be of use to speed up computations of marginals, for example, when the main source of partial distinguishability are perturbations to the polarization state of the photons.  

In more general scenarios though, the $n$ internal states of the photons span a Hilbert space of dimension at most $n$ and so this result is of limited use as $W(\eta)$ can have a rank which scales with the system size, implying that computing the marginals exactly with this method takes exponential time in the system size. In this case, we can may use different approaches allowing us to exploit partial distinguishability for more efficient approximations of the characteristic function. For example, using the techniques from Refs.~\cite{RenemaOxfordPaper, renema2020marginal}, we may obtain approximations where the error scales as $\log(1/x)$, where $x$ represents a distinguishability parameter.





\section{Signatures of multiphoton interference in Fourier interferometers}
\label{sec:Fourier}
In this section, we give analytical evidence that the photon distribution in binned output modes contains important information about multiphoton interference. To do so, we focus on the 
Fourier interferometer,  defined by the unitary transformation 
$$F_{jk} = \frac{1}{\sqrt{m}}e^{-2\pi i (j-1)(k-1)/m}.$$
This interferometer that has been widely studied in the context of validating multiphoton interference. Due to its symmetries, it has been shown that most of the outcome probabilities are suppressed (that is, equal zero) if the inputs are fully indistinguishable \cite{tichy2014stringent}. A violation of these suppression laws can be used to test indistinguishability of the input photons. 

Here we demonstrate that the formalism discussed in Sec.~\ref{sec:formalism} can be used to obtain analytical results about how photons distribute in subsets of output modes of a Fourier interferometer. We show that even considering a single subset, the way indistinguishable photons behave is drastically different than distinguishable ones. We focus on two interesting examples, namely, the computation of the single-mode density matrix and on the photon-counting distribution on the odd output modes. Our results go beyond suppression laws as they allow us to predict the full distribution in these subsets and not only which events are suppressed. 

\subsection{Single-mode density matrix}\label{sec:single_mode_Fourier}
One of the simplest ways of looking for signatures of multiphoton interference is by measuring subsystems of the output state of the linear interferometer, i.e.~the reduced state of a few output modes. We focus here on the single-mode density matrix of a Fourier interferometer, with a single-photon in each of the input modes, i.e.~$n=m$. This in an interesting setting as it falls within the scope of the results of Refs. \cite{cushenhudson1971, becker2021convergence}, which allows us to predict that the asymptotics of the single-mode density matrix is given by a thermal state with average photon number $\langle n \rangle = 1$.

To our knowledge, the exact form of the distribution for finite-sized systems has not been shown explicitly before. We show in Appendix \ref{secapp:single_mode_FT} that the formalism of Sec.~\ref{sec:formalism} allows us to obtain this distribution analytically. The probability of observing $k$ photons, if the input photons are fully indistinguishable, is given by
\begin{equation}
P_k^B= \sum_{a=k}^n (-1)^{k+a} {{a}\choose{k}} {{n}\choose{a}} \frac{a!}{n^a}. 
\end{equation}
Although this distribution is well approximated by the geometric distribution with a mean equal to 1, it has some important differences. For example, the probability of observing $n-1$ photons is always 0, a fact that can also be predicted from suppression laws \cite{tichy2010zero}. 
In contrast, interference of distinguishable photons results into a very different distribution. Using simple combinatorial arguments, one can see that the probability of observing $k$-photons in a single mode is given by the binomial distribution 
\begin{equation}
P_k^D= {{n}\choose{k} }\frac{1}{n^k}\left(1-\frac{1}{n}\right)^{n-k}.
\end{equation}
Asymptotically, this tends to a Poisson distribution with a mean equal to 1.  This shows that photon distinguishability already plays a significant role in the photocounting statistics of a single detector.

\subsection{Photon number distribution in larger subsets}\label{sec:special_subset}

\begin{figure}
\centering
\includegraphics[width=0.9\linewidth]{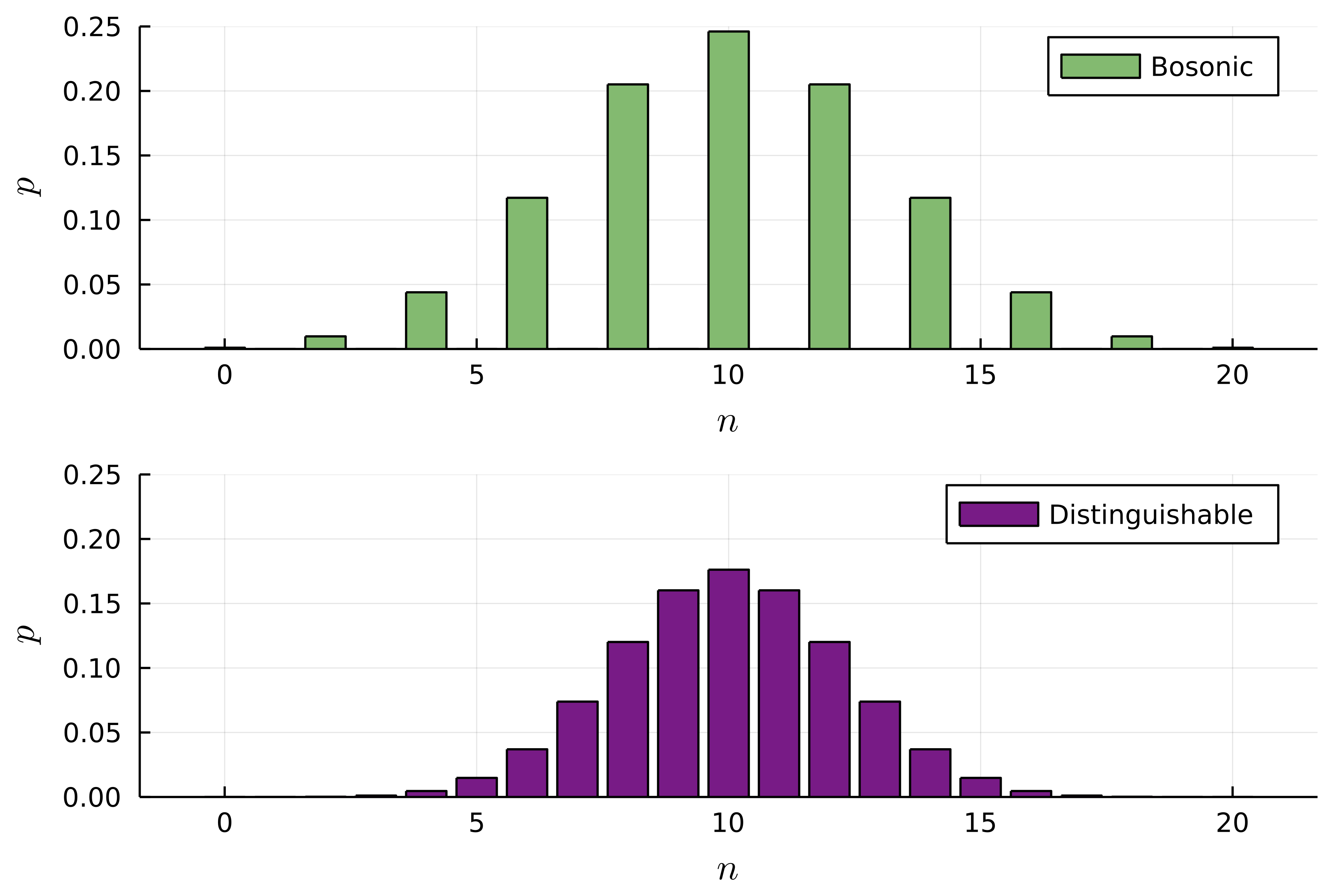}
\caption{Comparison of the probability of seeing $k$ photons in the odd output modes of the Fourier interferometer with $1$ particle per input mode. For bosons we see a suppression of events with odd $k$, whereas the events with even $k$ follow a binomial distribution. For distinguishable particles the probabilities follow a simple binomial distribution.  }
\label{fig:FourierAodd}
\end{figure} 
In the previous section, we have chosen to look at a very simple subset of the output modes, given by a single mode. Our formalism, however, allows us to look for signatures of multiphoton interference by considering more general subsets. If the interferometer has some particular symmetries, it is expected that choices of subsets that reflect these symmetries should reveal larger differences between the behavior of indistinguishable and distinguishable photons \cite{shchesnovich2017asymptotic}. We give a specific example in what follows. Let us consider again the Fourier interferometer with a single photon in each of the input modes and analyse how many photons end up in the odd modes, i.e.~the set $\setparti=\{1,3,\dots,n-1\}$, where we take the number of modes $n$ to be even. Using the formalism of Sec.~\ref{sec:formalism}, we compute analytically the photon-counting probabilities in this subset in the two extreme cases of indistinguishable vs distinguishable photons, which we plot in Fig.~\ref{fig:FourierAodd} (see Appendix~\ref{secapp:special_subset} for a detailed derivation). For ideal bosons, we obtain 
\begin{equation}\label{eq:bosonic_oddmodes}
P^B_k = \begin{cases}
0 &\mbox{if }k \mbox{ is odd},\\
\frac{1}{2^{n/2}}{n/2 \choose k}&\mbox{if }k \mbox{ is even}.
\end{cases}
\end{equation}
Events with an odd number of photons are fully suppressed, whereas the events with an even photon number follow a simple binomial distribution. A sharp contrast is observed with respect to the behavior of distinguishable photons, which follows a simple binomial distribution 
\begin{equation}
P_k^{D} = \frac{1}{2^n}{n \choose j}.
\end{equation}
This suggests that, in an experimental setting, the analysis of photon distributions in properly chosen subsets may be used to diagnose partial distinguishability in the input photons. In particular, it would be interesting to investigate whether the measured statistical deviations to the ideal distribution of Eq.~\eqref{eq:bosonic_oddmodes} may be used to bound the degree of genuine multiphoton indistinguishability of the input state \cite{brod2019witnessing}. 
\section{Validation of boson samplers}\label{sec:validation}
The main premise of our work is that, from the way photons distribute in partitions of the output modes of a boson sampler, it is possible to tell whether we are in the presence of an ideal boson sampler from a noisy one. In this section we justify this claim for Haar-random interferometers.  First, we introduce the models of noise we analyse, namely  partial distinguishability and photon loss. Subsequently we give analytical and numerical arguments showing the binned output distributions are sensitive to partial distinguishability and may be used for efficient validation tests. Finally, we stress that taking into account outcomes with a few lost photons may significantly speed-up validation tests as they still carry information about photon distinguihability. 
\subsection{Noise models}\label{subsec:noise_models}
Although the formalism of Sec.~\ref{sec:formalism} is able to encompass arbitrary inputs of partially distinguishable photons, here we consider a specific model of partial distinguishability, for the sake of performing numerical simulations about the validation method we propose in this work.  We assume that the wave-functions describing the internal degrees of freedom of any pair of photons have an overlap given by a distinguishability parameter $0\leq x \leq 1$. The off-diagonal elements of the distinguishability matrix $S$ from Eq.~\eqref{eq:S_matrix} are thus $S_{ij}=x$ whereas the diagonal elements are one by definition. In this model, the distinguishability matrix is a convex interpolation of the distinguishability matrices of two extreme cases. At $x = 0$, the photons behave as fully distinguishable particles and the observed statistics corresponds to the classical case (no interference),  when each one of the photons is sent at different times to the interferometer. At $x = 1$, we recover the ideal boson sampling case of linear interference between fully indistinguishable bosons. This interpolation model has been widely studied in works such as Refs.~\cite{tichy2015_partial_distinguishability, shchesnovich2015partial, shchesnovich2021distinguishing, RenemaOxfordPaper}, both from the perspective of understanding interference phenomena in the "quantum-to-classical" transition or with the aim of providing efficient classical simulation algorithms for noisy boson samplers. 

Regarding photon loss, we restrict to the uniform loss model (Fig.~\ref{fig:lost_photons}). Each photon has a probability $0 \leq \eta \leq 1$ to go through the interferometer and thus lead to a detection. Mathematically, this is equivalent to adding a beam-splitter of transmissivity $\eta$ at the end of each output source, with the reflected branch of the beam-splitter becoming an environment mode to which the photon can be sent (and thus represent a lost photon). 
We are interested in the photon configuration in a partition of the first $m$ modes, from which we can infer how many photons where lost to the environment modes. 

\subsection{Comparison to asymptotic formulas}
The results in Sec.~\ref{sec:Fourier} indicate that the photon-counting statistics in output mode partitions can reveal striking signatures of multiparticle interference in certain symmetric interferometers. It is important to understand if this is also true in the usual boson sampling scenario, where the unitary characterizing the interferometer is drawn at random from the Haar measure. This question was addressed in the work of Shchesnovich in Refs.~\cite{shchesnovich2017quantum, shchesnovich2017asymptotic}, with the derivation of asymptotic laws characterizing how photons distribute in the binned output modes in large interferometers.  Using combinatorial arguments, it was found that these photon-counting probabilities, when averaged over the Haar-random interferometers, are given by
\begin{align}
    \label{eq:asymptoticsD}
    p^D(\vect{k}) &= \frac{n!}{\prod_{z=1}^{K} k_z!}\prod_{z=1}^{K} q_z^{k_z} \\
    \label{eq:asymptoticsB}
    p^B(\vect{k}) &= p^D(\vect{k}) \frac{\prod_{z=1}^{K} (\prod_{l=0}^{k_z-1} [1+l/K_z]
    )}{\prod_{l=0}^{n-1} [1+l/m]}
\end{align}
where $D$ signifies distinguishable particles, and $B$ fully indistinguishable (bosonic) ones. We also define the bin size $K_z=|\setparti_z|$ as well as the relative bin size $q_z = K_z/m$. Here, it is assumed that the $K$ bins span all the output modes, so from particle number conservation we have that $k_{K}=n-\sum_{z=1}^{K-1}k_z$. Assuming a constant relative bin size $q_z$, the  asymptotic form of the previous expressions, as $n,m \rightarrow \infty$, is given by a multivariate Gaussian:  
\begin{align}
    \label{eq:haarAveragePartitions}
    P^{\sigma}(\vec{k}|K) &= \frac{\exp \{ 
    -n \sum_{z = 1} ^K
    \frac{(x_z-q_z)^2}{2(1+\sigma \rho)q_z}
    \}}{(2\pi(1+\sigma \rho)n)^{(K-1)/2} \prod_{z = 1} ^K \sqrt{q_z}} \nonumber \\
    &\times \left(
    1 + \mathcal{O}
    \left( \frac{\rho \delta_{\sigma, +}}{n}
    \right)
    \right),
\end{align}
with $x_z = k_z/m$ and $\sigma = 1$ for indistinguishable particles and $0$ for distinguishable ones. The difference between the behavior of these two extreme cases shows up via the particle density $\rho=n/m$ which influences the standard deviation of the Gaussian statistics when the input particles are ideal bosons. For the technical details about the validity regime of the asymptotic formula, as well as the error of this approximation, we refer to Ref. \cite{shchesnovich2017asymptotic}. 

\begin{figure}
  \centering
  \includegraphics[width=0.45\textwidth]{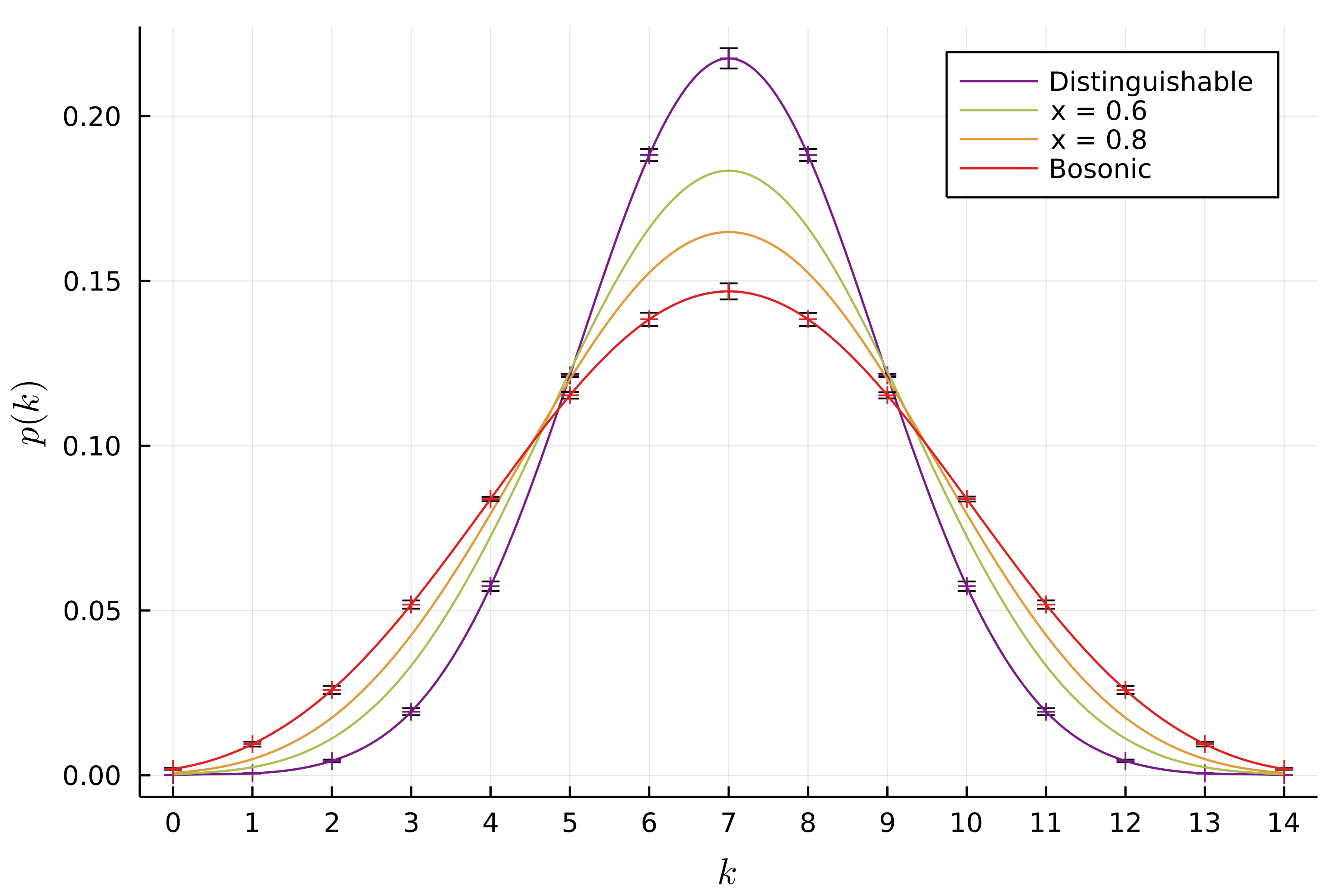}
  \caption{\textbf{Effect of partial distinguishability on a subset distribution.} We consider the photon number distribution in a subset consisting of the first half of the $m=14$ output modes with $n=14$. The probabilities are averaged on $1000$ Haar-random unitaries. Error bars show one standard deviation. We see that they are well approximated by a Gaussian distribution, with the main difference between each case being the width of the curve.}
  \label{fig:bosonic_to_distinguishable}
\end{figure}

 These results suggest that the probability distribution in binned output modes is sensitive to partial distinguishability between the photons even for Haar-random unitaries. To our knowledge, there exist no explicit asymptotic formulae in this scenario 
 and so we resort to numerical simulations to confirm this hypothesis, using the method detailed in Sec.~\ref{sec:formalism}. 
 For simplicity, we consider a bipartition of the output modes into two sets of equal size and the partially distinguishability model introduced in Sec.~\ref{subsec:noise_models}, which interpolates between distinguishable and indistinguishable particles via a indistinguishability parameter $x$. 
 The observed distribution for 14 photons in 14 output modes for several values of $x$ is plotted in Fig.~\ref{fig:bosonic_to_distinguishable}. 
 The figure reveals significant differences in the probabilities as the indistinguishability parameter is varied.  The width of the bell-shaped curve decreases as photons become more and more distinguishable, as suggested by the asymptotic formulas from Eqs.~\eqref{eq:asymptoticsD} and \eqref{eq:asymptoticsB} which describe the extremes. 
 This indicates that boson bunching effects play a role, since events where a large fraction of the photons are observed in the same bin are more likely as the photons become more indistinguishable.   


\subsection{Distance between distributions}\label{subsec:distance_distributions}
 A standard quantity used to quantify the distance between probability distributions is the total variation distance (TVD), defined as 
 \begin{equation}
     \tvd{p,q}=\sum_j |p_j-q_j|.
 \end{equation}
 This is an especially pertinent metric regarding the problem of distinguishing two distributions \cite{valiant2017automatic, blais2016alice} via sampling, since the number of samples $n_s$ needed to distinguish a distribution $p$ from another $q$ scales as 
\begin{equation}\label{eq:number_samples_TVD}
    n_s = \mathcal{O}(\tvd{p,q}^{-2}). 
\end{equation}
In what follows, we analyse how the TVD varies in different cases, depending on system size, partial distinguishability and loss. 

For the rest of this paper, and unless specified otherwise, we will always select the bins to form an equipartition, as defined in Appendix \ref{appendix:equipartition}.

\begin{figure}
  \centering
     \includegraphics[width=0.5\textwidth]{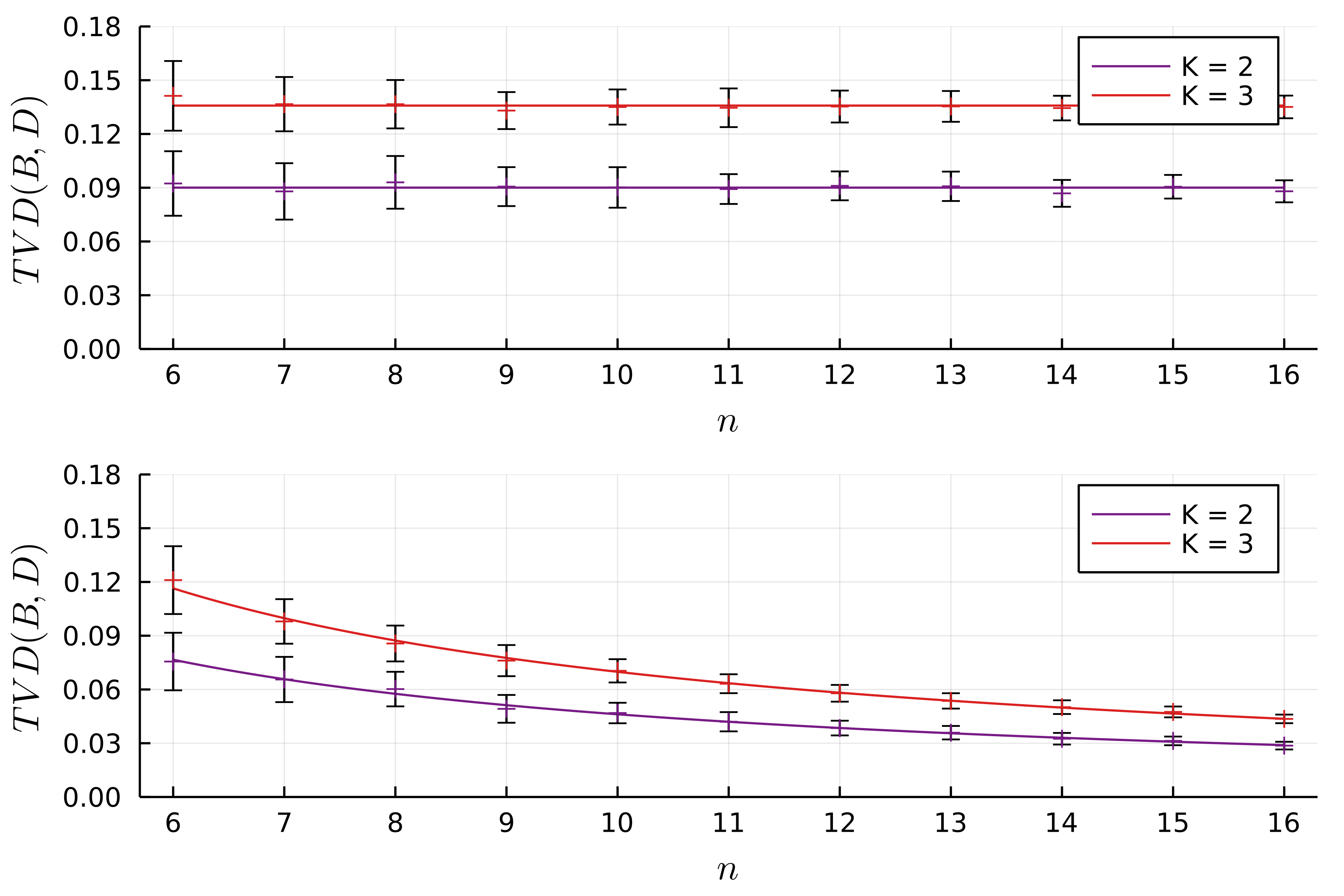}
  \caption{\textbf{Evolution of the TVD with the system size.} Haar-averaged TVD between the binned output distributions of bosonic and distinguishable inputs with varying system size. We consider two equipartions of size $K=2$ and $K=3$. In the top subplot, we consider a sparse but constant density $m = 5n$, while in the bottom subplot it is the no-collision regime $m = n^2$. The error bars represent the standard deviation when averaged over $100$ trials. Fluctuations come from the intrinsic random Haar sampling, as well as the limited batch size used in the averaging. It can be seen that coefficient of variation decreases with increased system size (see Appendix~\ref{appendix:numerics}). We also see a significant increase of the TVD when we choose a larger partition size.}
  \label{fig:size}
\end{figure}

\paragraph{Bosons vs. distinguishable particles} 
First, let us compare the two extreme cases of indistinguishable vs distinguishable particles.  
In Fig.~\ref{fig:size}, we analyse how the TVD, averaged over Haar-random unitaries, depends on the number of input photons. We do so in two different scenarios: when the density $\rho=n/m$ is constant as well as in the regime usually considered in boson sampling, where the number of modes $m=O(n^2)$ (and thus $\rho=1/n$), which ensures that the probability of observing events with collisions is small \cite{bosonsampling}. As suggested by the asymptotic formulae, the density plays an important role. For constant density the TVD remains constant independently of the number of photons and consequently, the number of samples needed to distinguish the two distributions does not scale with the system size. In contrast, the bottom curve in Fig.~\ref{fig:size} suggests an inverse polynomial decay for the TVD in the \emph{collision-free} regime. This implies that the two distributions can still be distinguished efficiently, i.e.~with a polynomial number of samples. We also remark the significant increase of the TVD if we take a larger partition size.
For example, we observe that the TVD roughly doubles when we compare $K=2$ with $K=4$, which implies we need 4 times less samples to distinguish the two distributions, according to Eq.~\eqref{eq:number_samples_TVD}.

\begin{figure}
  \centering
  \includegraphics[width=0.5\textwidth]{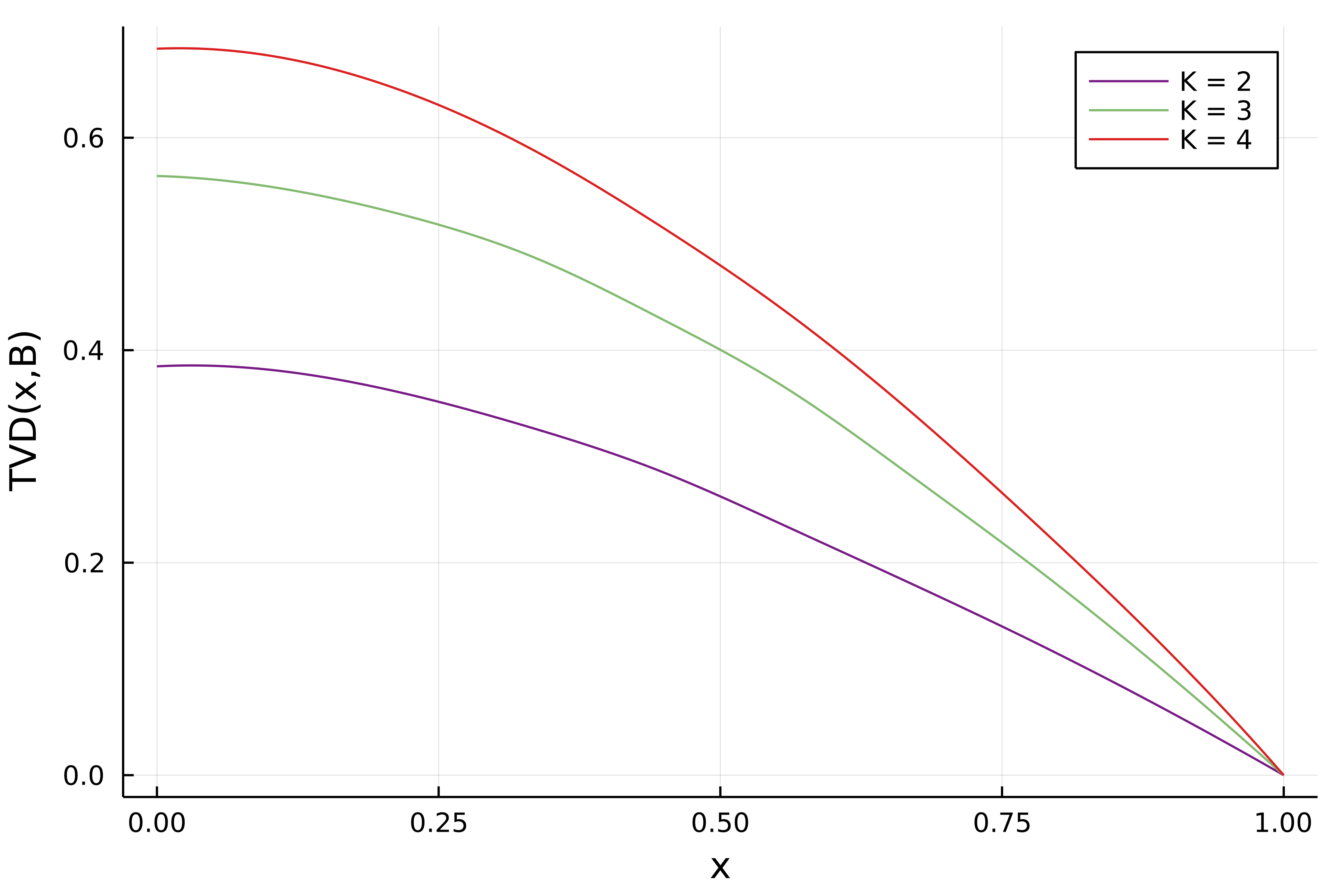}
  \caption{\textbf{Effect of partial distinguishability on TVD.} The TVD between the binned distributions corresponding to ideal photons ($x = 1$) and partially distinguishable ones is displayed as a function of the distinguishability parameter $x$. In this figure we took $m=n=10$ and equipartitions of size $K = 2,3,4$. }
  \label{fig:x_model}
\end{figure}

\paragraph{Partial distinguishability}
As previously mentioned, partial distinguishability between the input photons is one of the main sources of noise in boson samplers and may render the experiment easy to simulate classically \cite{RenemaOxfordPaper, moylett2019classically}. Hence, a good validation test should be able to differentiate between an ideal boson sampler from one with partially distinguishable photons. To have a better understanding about the sensitivity of our validation test to partial distinguishability we again make use of the the one-parameter model interpolating between distinguishable and indistinguishable photons (see Sec.~\ref{subsec:noise_models}). In Fig.~\ref{fig:x_model} we compare the distance between the photon-counting probabilities in the partitions when the input photons are indistinguishable ($x=1$) and when they are partially distinguishable, as a function of the parameter $x$. We see a sharp increase in the TVD as we move away from the ideal case, suggesting that the probability distributions can be distinguished in practical scenarios. 

Similarly, we have also analysed how the variation of the TVD between the distributions coming from ideal bosons or partially distinguishable ones (with some fixed $x$) varies as a function of the system size. Interestingly, the behavior follows the same trend as that that of Fig. \ref{fig:size}: for constant densities the TVD remains constant whereas in the ``collision free" regime it suggests a polynomial decay. A specific example for $x = 0.9$, can be found in Fig. \ref{fig:sizeWithX} of Appendix \ref{appendix:numerics}. This numerical evidence strongly suggests that the method of analysing photon-counting distributions in subsets can efficiently distinguish ideal boson samplers from ones with partially distinguishable inputs.  

\paragraph{Dependency on photon density}
The previous results reaffirm the important role of photon density in the efficiency of discriminating ideal and noisy boson samplers. Although analytical results about this dependency may be difficult to obtain, we may use the numerical data to extract power laws that approximately govern this behavior. For different values of partial distinguishability, and considering equipartitions with a small number of subsets, the data suggests that TVD between the ideal distribution and that coming from partially distinguishable input photons with a fixed $x$ is approximately described by the following behavior
\begin{equation}
    \tvd{B, x} \simeq c(K,x)\rho.
\end{equation}
Here, $\rho=n/m$ is the photon density and $c(K,x)$ is a numerical constant depending on the number of subsets $K$ and the level of partial distinguishability $x$. More precisely, when fitting the numerical data with an ansatz model $\tvd{B, x} \simeq c(K,x)\rho^r$, we obtain a value of $r\approx 0.95$, with some small variability depending on the value of partial distinguishability chosen, which may be also due to the finite number of trials. Further plots and details regarding the quality of the approximation are given in Appendix \ref{appendix:numerics}.  While not formally proven, this approximate power law in the regimes we explored further suggests the efficiency of the validation scheme, with a polynomial decrease of the TVD between binned distributions when $\rho$ decreases polynomially in the number of photons.

\subsection{Hypothesis testing}\label{subsec:hyp_testing}
Given a collection of experimental samples and two possible theoretical descriptions of the experiment, the formalism of Bayesian hypothesis testing allows us to predict how many samples are needed to decide which one is more likely to describe the observed data. This strategy has been exploited in the context of boson sampling in Refs.~\cite{bentivegna2015bayesian,dai2020bayesian,flamini2020validating}. We may assume that one of the hypothesis to describe the experiment is an ideal boson sampler, with indistinguishable input photons. We call this the null hypothesis $H_0$, which we would like to test against an alternative description of the experiment $H_a$. The later could be for example a boson sampler with distinguishable or partially distinguishable input photons. Given an output sample $\vect{s}= (s_1, s_2, \dots, s_m)$, we can compute the ratio between the probability of observing this sample assuming the null hypothesis $p(\vect{s} | H_0)$, and its counterpart assuming the alternative hypothesis $p(\vect{s} | H_a)$. The product of these ratios over the different samples give us the Bayesian factor 
\begin{equation}
    \label{eq:bayesian}
    \chi = \prod _{i=1}^{n_s} \frac{p(\vect{s}^{(i)} | H_0)}{p(\vect{s}^{(i)} | H_a)}, 
\end{equation} 
where $\vect{s}^{(i)}$ refers to the $i$-th sample and $n_s$ to the total number of samples. The confidence in the hypothesis $H_0$ can be computed from $\chi$ as 
\begin{equation}
    p_{null} = \chi/(\chi+1). 
\end{equation}
Although there is numerical evidence that this method requires only a modest number of samples to validate an ideal boson sampler against certain alternative hypothesis~\cite{bentivegna2015bayesian,dai2020bayesian,flamini2020validating}, the main drawback is that the computation of the confidence $p_{null}$ is not efficient. Indeed, the output probabilities of ideal boson samplers $p(\vect{s}^{(i)} | H_0)$ are exponentially hard to approximate. 

\begin{figure}
  \centering
  \includegraphics[width=0.5\textwidth]{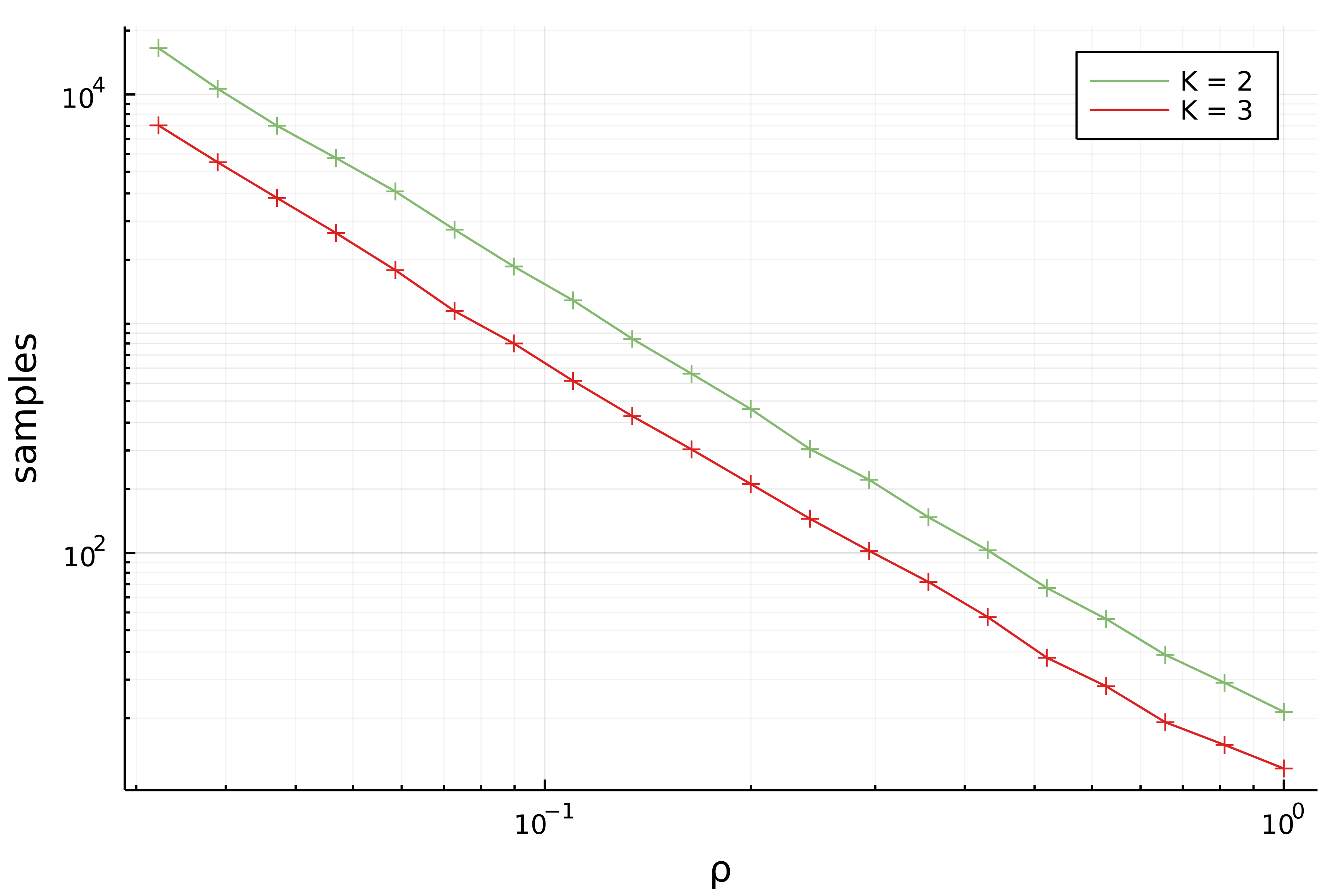}
  \caption{\textbf{Number of samples} required to distinguish the indistinguishable and distinguishable inputs. 
  This number of samples is determined by taking the average number of samples required to obtain a certainty of $p_{null} = 95\%$ using the Bayesian approach laid out in the main text, see \eqref{eq:bayesian}.
  We numerically confirmed that this quantity depends to a good approximation only from boson density $\rho$ in numerical trials, as is discussed in the main text for the TVD. This specific figure was generated using parameters $n=10$, $m=10,\dots,300$ and each point is an average over $1000$ Haar-random unitaries.}
  \label{fig:number_samples}
\end{figure}

\begin{figure}
  \centering
  \includegraphics[width=0.5\textwidth]{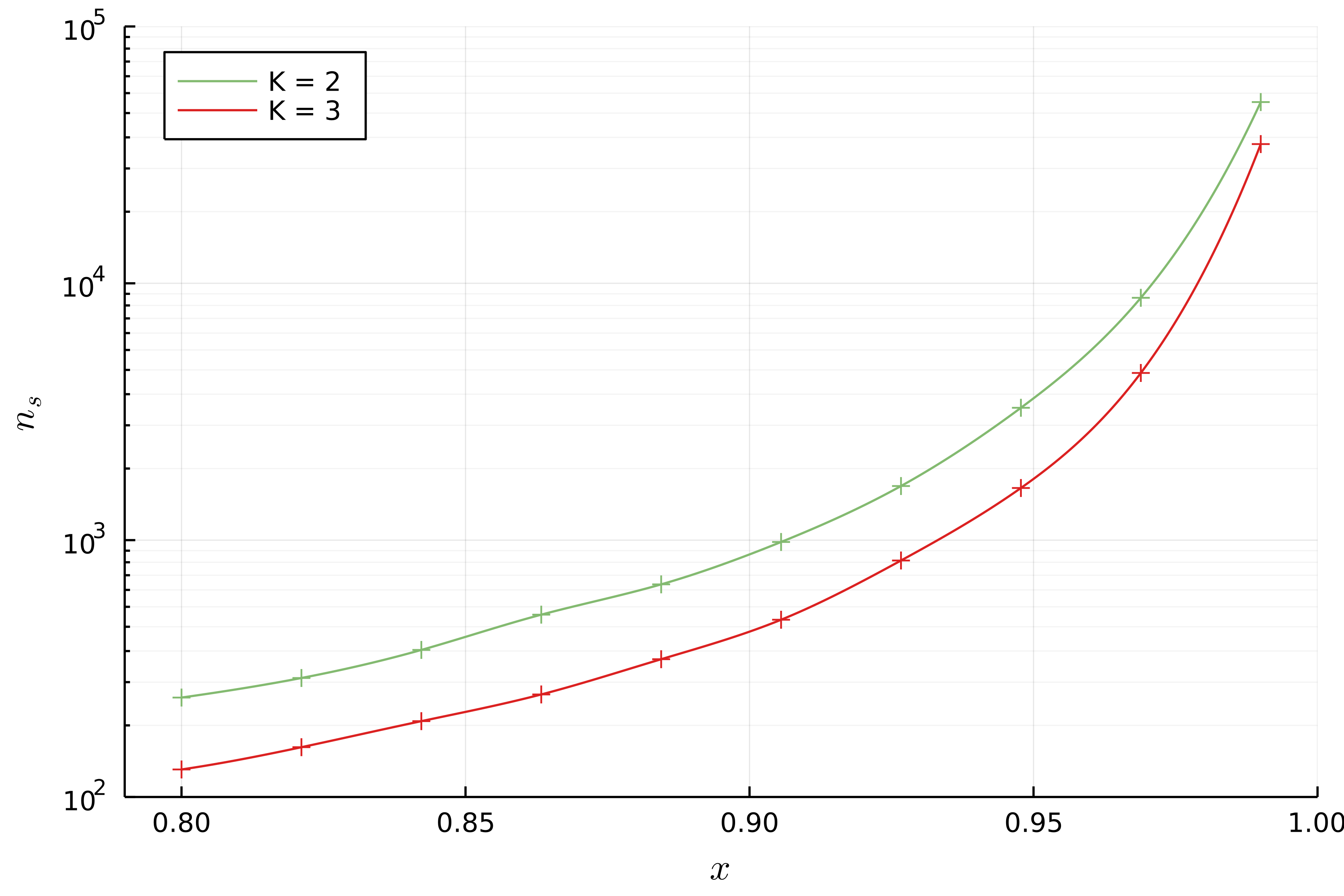}
  \caption{\textbf{Number of samples} required to reject the null hypothesis of having indistinguishable bosons when the input is actually the one parameter interpolation with indistinguishability $0.8\leq x \leq 0.99$. 
  This number of samples is determined by taking the average number of samples required to obtain a certainty less than $p_{null} = 5\%$ using the Bayesian approach laid out in the main text, see \eqref{eq:bayesian}.
  This specific figure was generated using parameters $n=10$, $m=10$ and each point is an average over $1000$ Haar-random unitaries.}
  \label{fig:number_samples_x}
\end{figure}

In this section, we consider as output samples the events $\vect{k}$, corresponding to the photon number distribution in $K$ bins as discussed in Sec.~\ref{sec:formalism}. We show that this simpler-to-compute probability distribution can be used to validate boson sampling experiments, instead of the full outcome distribution. In particular, we are interested in the following question: how many samples do we need to reject the hypothesis that we have an ideal boson sampler when we are in the presence of a noisy one? To give an example, we again focus on the interpolating model of partial distinguishability discussed in Sec.~\ref{subsec:noise_models}. In Fig.~\ref{fig:number_samples_x}, we plot the number of samples needed to reject the null hypothesis with a confidence of $95\%$ if the experiment is described by a boson sampler whose input has a distinguishability parameter $x$. We observe that for 10 photons in 10 modes and a distinguishability parameter $x=0.8$, a few hundred samples are enough to reject the null hypothesis even in the simplest case where we choose two equal-sized bins. This is improved by a factor of about one half if we bin the output modes into three subsets, thus gaining more information about the full probability distribution. We remark also that, as expected, the number of samples to reject the null hypothesis sharply increases as the noisy boson sampler becomes closer to ideal, i.e.~as $x$ tends to one. 

Numerical evidence also suggests that it is possible to extract approximate power laws, that allow us to predict the number of samples needed as a function of the photon density, in analogy to what was done for the TVD in Sec.~\ref{subsec:distance_distributions}. In the case where the task is to differentiate between  distinguishable vs. indistinguishable input photons, we verify numerically that this dependence is well described by the following power law  
\begin{equation}
    \label{eq:number_samples}
    n_s \approx \frac{d(K)}{\rho^{5/2}},  
\end{equation}
which we extract from the data of Fig.~\ref{fig:number_samples}. Here, $d(K)$ a constant depending on the choice of partition and level of partial distinguishability. Similar power laws may be extracted when the input photons are partially distinguishable (more details in Appendix \ref{appendix:numerics}). Such extrapolations are useful to predict the necessary sampling rates for validate experiments when scaling up the system size.  

\subsection{Validation in the presence of loss}
Thus far, we have not yet considered the role of loss in the validation task.  The average number of photons that go through the linear optical circuit decreases exponentially with the circuit depth \cite{garcia2019simulating}, which is usually linear with the number of modes. Hence, most of the experimental observations will be of lossy events and even though postselection on "lossless" events is possible, it would lead to an exponential decrease of the sampling rate. For this reason, it is interesting to consider events with a few lost photons for the task of validating an experiment trying to demonstrate a quantum computational advantage -- this not only increases significantly the sampling rate but also these experiments may still be difficult to simulate with classical algorithms, provided that the photons are fully indistinguishable from each other \cite{aaronson2016bosonsampling}. 

\begin{figure}
  \centering
  \includegraphics[width=0.5\textwidth]{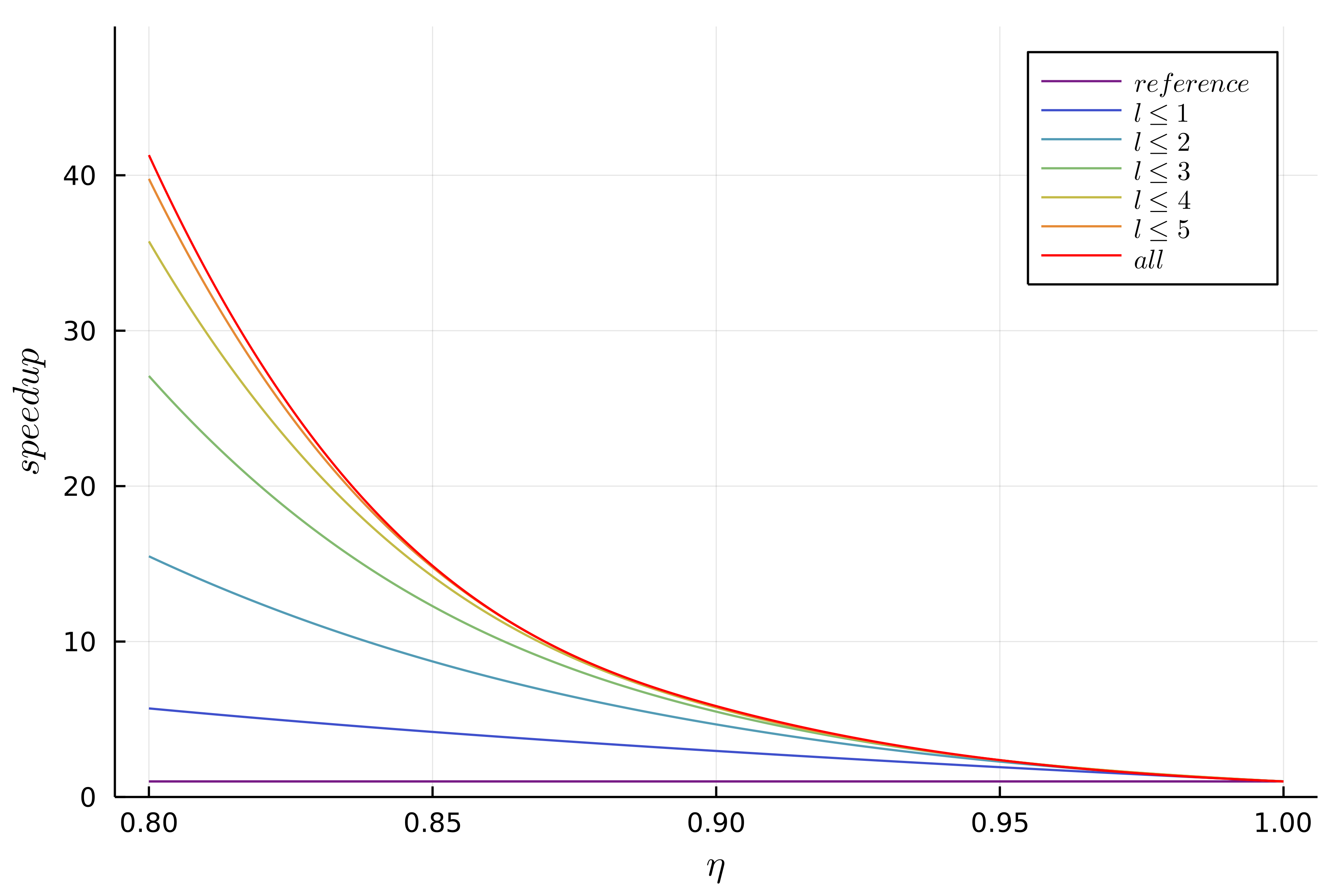}
  
  \caption{\textbf{Validation speed up using lost photons.} 
  In this figure, we take a single subset of half the $m = 10$ output modes with $n=10$ photons. 
  We validate that the observations are indistinguishable bosons versus the one-parameter model with $x = 0.9$.
  We consider a model of uniform loss, with single photon transmissivity $\eta$. 
  The reference curve shows validation with no lost photons (y-coordinate is $1$).
  Curves with $l \leq 1,\dots,5$ show that the including output data with up to $l$ photons lost gives significantly a faster validation scheme. This speed up becomes less and less important as $l$ increases. 
  The data is averaged over $100$ Haar-random unitaries and $1000$ validation runs for each unitary.}
  \label{fig:lost_photons}
\end{figure}

The question we address in this section is whether such lossy events can be used to validate the experiment faster, i.e.~whether they contain useful information about other sources of noise affecting the experiment, namely photon distinguishability, which may render the experiment easy to simulate classically.  
Let us consider again the hypothesis testing setting using the data from how photons distribute in subsets of output modes. Here we consider only a single subset with half the modes for simplicity. 
We define the validation time as the time needed to distinguish between the ideal hypothesis $H_0$ and an alternative one $H_a$ with some predetermined confidence (say 95\%), assuming a constant sampling rate of the lossy boson sampler. 
As we will see, using data with lost photons to test for photon distinguishability, may lead to a significant speed-up in the validation time. 
To give a concrete example, we consider the task of validating a boson sampler with 10 photons in 10 modes for different loss parameters. We assume our boson sampler has a partially distinguishable input state with distinguishability parameter $x=0.9$ (hypothesis $H_a$) and the task is to test if it is an ideal one with $x=1$ (hypothesis $H_0$). 
We define $T_l$ as the average validation time over Haar-random interferometers, if we take into account the data up to $l$ lost photons. 
In Fig.~\ref{fig:lost_photons}, we plot the ratio $T_0/T_l$ as a function of the loss rate which we assume to be uniform. This quantity reflects the average speed-up obtained by considering data with lost photons.
For a loss rate of 0.2, a speed-up of around 40-fold is obtained when considering all the data, independently of how many photons were lost. Fig.~\ref{fig:lost_photons} also reveals that, as expected, events where more photons were lost contain less information about the distinguishability of the input. This is visible, for example, from the fact that the speed-up obtained when taking the data with up to five lost photons is very similar to taking all the data.   
We have verified numerically that this speed up tends to increase in larger systems, even for a fixed $\eta$.

\section{Discussion}
In this work, we have showed that a coarse-graining of the boson sampling output distribution by grouping the output modes into bins, provides a simpler to analyse outcome distribution and a natural validation test for boson samplers. We demonstrate that, given a theoretical model of the experiment, the binned output distribution can be classically approximated as efficiently as if we run the experiment itself. The main technique we use to obtain this result is the computation of the (discrete) characteristic function of the binned distribution via Gurvits randomized algorithm for permanent approximation \cite{bosonsampling}.

Even for a small number of output bins, the distribution reveals great sensitivity to photon distinguishability. Our numerical simulations using this validation test suggest that a polynomial number of samples is sufficient to distinguish an ideal boson sampler from one with partially distinguishable input photons. We also showed how in realistic situations -- where the experimental data includes a vast majority of events where photons are lost -- the outcomes with lost photons contain useful information about partial distinguishability and that the effective use of this data can greatly speed-up validation tests.     

Multiple interesting research questions arise related to this validation test. If we do not trust that the data is coming from an actual physical experiment, can we guarantee that no efficient classical algorithm exists that may spoof the test? An important property of the validation test we consider in our work is that the subsets we choose to test the experimental data can be chosen arbitrarily \emph{a posteriori}. At first sight, this makes the test harder to spoof: a potential adversary trying to mimic the behavior of an ideal boson sampler would have to generate samples such that they are consistent with the correct coarse-grained photon-counting distributions for exponentially many possible subset choices. However, one may wonder whether the knowledge of the analytic form of the average over Haar-random unitaries of the binned-output distribution (see Eq.~\eqref{eq:asymptoticsB} and Refs.~\cite{shchesnovich2017asymptotic, shchesnovich2017quantum}) may be used to spoof the test. We argue in Appendix \ref{appendix:spoofing} that the knowledge of the mean of the binned distributions is not enough to spoof the test, because different partitions choices exhibit significant fluctuations around the mean which should be measurable with a polynomial number of samples.

Another interesting question is whether it is possible to obtain analytical results corroborating our numerical evidence about the sample efficiency of the method. Previous results on validation tests based on generalized bunching probabilities from Ref.~\cite{shchesnovich2016universality}, which can be seen as a particular outcome of a binned distribution, suggest that comparing bosonic and distinguishable particles can be done in a sample efficient way. However, the problem becomes more difficult when different models of partial distinguishability come into play. 

Moreover, while we set our interest in using binned output probabilities as a validation method, we also believe that it could be a useful tool for probing partial distinguishability. Deviations from the expected binned output distributions may possibly be used to quantify the degree of indistinguishability of the input photons, specially in highly symmetric interferometers such as the Fourier transform, where large differences between distinguishable and indistinguishable particles are observed.

Our work also opens up the question of whether certain decision or function problems that can be solved by boson samplers proposed in Refs.~\cite{nikolopoulos2016decision, nikolopoulos2019cryptographic}, with potential cryptographic applications, may actually be solved by efficient classical algorithms. Some of these problems are also based on questions related to probability distributions obtained after certain binning of the boson sampling data. Even though the binning procedure is not directly equivalent to ours, it would be worth investigating if the binned distributions from Refs.~\cite{nikolopoulos2016decision, nikolopoulos2019cryptographic} may be approximated via a similar formalism to that presented in Sec.~\ref{sec:formalism}.  

During the completion of this work, we became aware of the recent works from Refs.~\cite{oh2022quantum, dellios2022validation, drummond2022simulating}. Ref.~\cite{oh2022quantum} develops an efficient classical algorithm to approximate molecular vibronic spectra, using a formalism similar to Sec.~\ref{sec:formalism} based on approximating Fourier components of the target probability distribution using Gurvits algorithms. In turn, Refs.~\cite{dellios2022validation, drummond2022simulating} consider photon-number distributions in binned output modes as validation tests of Gaussian boson samplers. The authors use phase space methods to approximate these distributions, which are referred to as group-count probabilities. In contrast, we focus on validation of standard boson samplers and develop a different formalism to compute group-count probabilities which does not involve phase space averages. Another difference of our work is that, while the authors of Refs.~\cite{dellios2022validation, drummond2022simulating} focus on noise sources more likely to affect Gaussian boson samplers, we focus on testing the sensitivity of the method to photon distinguishability as this is one of the main noise sources affecting standard boson samplers. Overall, we believe our contribution, together with those previous works, suggests that analysing how photons distribute in binned output modes is a scalable and practical method to use for validation of near-future experiments.       

Note that, for the sake of conciseness and ease of reading, we made the choice to limit the numerical analysis exposed in this paper to simple noise models such as uniform partial distinguishability and loss. In Ref.~\cite{seron2022bosonsamplingjl}, we provide tools that allow for general noise models, see below.





\section*{Code availability}

\label{sec:code_availability}

A complete Julia package, \href{https://github.com/benoitseron/BosonSampling.jl}{\textsc{BosonSampling.jl}}, and its related package, \href{https://github.com/benoitseron/Permanents.jl}{\textsc{Permanents.jl}}, includes
all the tools presented in this paper and many more regarding boson sampling. 
They are written in a user-friendly way and are aimed at experimentalists wanting to use this work. This package is already being used in boson sampling experiments.
The package is also focused on making it easy to write new models (such as noisy detectors, or new types of boson sampling) in an easy to write manner while being as fast as low-level languages such as C.

A related publication \cite{seron2022bosonsamplingjl} regarding \href{https://github.com/benoitseron/BosonSampling.jl}{\textsc{BosonSampling.jl}} is available on the arxiv.

A complete \href{https://benoitseron.github.io/BosonSampling.jl/stable/}{tutorial and documentation} are provided and interested users are welcome to contact the authors for possible extensions or specific needs.

All available Figures and data found in this article can be reproduced directly from the \verb|/docs/publication/partition/| folder of the package.

\section*{Acknowledgments}

The authors would like to thank F. Flamini, P. Valiant, G. Valiant for valuable discussions. 
B.S. is a Research Fellow of the Fonds de la Recherche Scientifique – FNRS. L.N. acknowledges support from the Fonds de la Recherche Scientifique – FNRS,   from FCT-Fundação para a Ciência e a Tecnologia (Portugal)
via the Project No. CEECINST/00062/2018, and from the European Union’s Horizon 2020 research and innovation program through the FET project PHOQUSING (“PHOtonic QUantum SamplING machine” - Grant Agreement No. 899544). N.J.C. acknowledges support from the Fonds de la Recherche Scientifique – FNRS under Grant No T.0224.18 and from the European Union’s Horizon 2020 research and innovation programme under the Marie Skłodowska-Curie grant
agreement No 956071. 
\\

\clearpage
\bibliographystyle{unsrtnat}
\bibliography{references_new_doi}

\clearpage
\onecolumn
\appendix

\onecolumngrid
\section{Transition amplitudes in the partially distinguishable case}\label{appendix:amplitudes_partially_distinguishable}

\subsection{Expression of the characteristic function as an amplitude}

Let us first show that the characteristic function introduced in Eq. \ref{eq:characteristicFunctionDef}
\begin{align}
    x({\vect{\eta}})&= 
    \E_{\vect{k}}\left[ \exp\left(i  \vect{\eta}\cdot \vect{k}\right)\right]\\
    & = \sum_{\vect{k}\in \Omega^{K}}P({\vect{k}}) \exp\left( i \vect{\eta}\cdot \vect{k}\right) \label{eq:expansion_xEtaAppendix}
\end{align}
can be expressed as through computing expectation values (also referred to as amplitudes)
\begin{align}
    \label{eq:xEtaAppendix}
    x({\vect{\eta}}) &= \braAket{\Psi_{out}}{e^{i\sum_{z = 1}^K\eta_z \op{N}_{\setparti_z}}}{\Psi_{out}}. 
\end{align}
Given an orthonormal basis for the internal states of the photons $\{ \ket{\Phi_j}\}$, we can expand any $n$-photon state with mode occupation numbers $\vect{s}=(s_1, \dots ,s_m )$ into the following orthonormal basis: 
\begin{equation}
    \ket{\vect{d}(\vect{s}), \Phi_{\vect{a}}}= \frac{1}{\sqrt{\mu(\vect{s})}}\prod _{j=1}^n \op{b}_{d_j(\vect{s}), \Phi_{a_j}}^\dagger\ket{0}.
\end{equation}
Here, $\vect{d}(\vect{s})$ is the  mode assignment list, a vector of dimension $n$ constructed as $\vect{d}(\vect{s}) = \oplus _{j = 1}^n \oplus _{k = 1}^{s_j} (j)$ \cite{tichy_tutorial}. The component $d_j(\vect{s})$ reflects the spatial mode occupied by the $j$th particle and is formally constructed by repeating $s_j$ times the mode number $j$. In turn, $\vect{a}$ is also a vector of dimension $n$, whose indices $a_j$ define that the internal state of the $j$th particle is $\ket{\Phi_{a_j}}$. Moreover, we also define $\mu(\vect{s})=s_1! \cdots s_m!$. 

The output state of a boson sampler with partially distinguishable input photons can then be written as 
\begin{equation}
\ket{\Psi_{out}} = \sum _{\vect{d}(\vect{s}), \vect{a}} \alpha(\vect{d}(\vect{s}), \Phi_{\vect{a}}) \ket{\vect{d}(\vect{s}), \Phi_{\vect{a}}}
\end{equation}
where 
\begin{align}
    \alpha(\vect{d}(\vect{s}), \Phi_{\vect{a}}) &= \braket{\prod _{j=1}^n \op{b}_{d_j(\vect{s}), \Phi_{a_j}}}{\Psi_{out}}
\end{align}
We can now expand part of the right side of Eq.~\eqref{eq:xEtaAppendix} as
\begin{align}
    &e^{i\sum_{z = 1}^K \eta_z \op{N}_{\setparti_z}}\ket{\Psi_{out}} \\
    &=e^{i\sum_{z = 1}^K \eta_z \sum _{j \in \setparti_z} \op{n}_j}\ket{\Psi_{out}} \\
    &=
    \sum _{\vect{d}(\vect{s}), \vect{a}} e^{i\sum_{z = 1}^K \eta_z \sum _{j \in \setparti_z} \op{n}_j} \alpha(\vect{d}(\vect{s}), \Phi_{\vect{a}}) \ket{\vect{d}(\vect{s}), \Phi_{\vect{a}}} \\
    &= \sum _{k_1=0}^n \dots \sum _{k_K=0}^n e^{i \sum_{z = 1}^K \eta_z k_z}
    \sum _{\vect{d}(\vect{s})|\vect{k}, \vect{a}} \alpha(\vect{d}(\vect{s}), \Phi_{\vect{a}}) \ket{\vect{d}(\vect{s}), \Phi_{\vect{a}}}
\end{align}
where $\vect{d}(\vect{s})|\vect{k}$ is understood as the vectors $\vect{s}$ compatible with finding the partition output count $\vect{k}$. By applying $\bra{\Psi_{out}}$, we obtain
\begin{equation}
    x(\vect{\eta}) = \sum _{k_1=0}^n \dots \sum _{k_\sizeparti=0}^n e^{i \sum_{z = 1}^\sizeparti \eta_z k_z} \sum _{\vect{d}(\vect{s})|\vect{k}, \vect{a}} |\alpha(\vect{d}(\vect{s}), \Phi_{\vect{a}})|^2
\end{equation}
Remark that  
\begin{equation}
    P(\vect{k}) = \sum _{\vect{d}(\vect{s})|\vect{k}, \vect{a}} |\alpha(\vect{d}(\vect{s}), \Phi_{\vect{a}})|^2
\end{equation}
is the probability to find the photon count $\vect{k}$, by construction, which proves that we recover Eq.~\eqref{eq:expansion_xEtaAppendix}. Therefore, by simply applying a $\sizeparti$-dimensional Fourier transform, we can obtain the binned output probabilities from the values of the $x_{\vect{\eta}}$'s by computing:
\begin{equation}
    \label{eqapp:Pk_FT}
    P(\vect{k}) = \frac{1}{(n+1)^\sizeparti}\sum_{l_1 = 0}^n \dots \sum_{l_K = 0}^n x\left(\frac{2\pi\vect{l}}{n+1}\right)e^{-2\pi i\vect{l}\cdot \vect{k}/(n+1)}
\end{equation}

\subsection{Computation of the amplitudes as permanents}

Let us derive the expression for $x({\vect{\eta}})$ from Eq.~\eqref{eq:xl_perm}. We recall that
\begin{align}
    x({\vect{\eta}})&= \bra{\Psi_\mathrm{out}}e^{i \vect{\eta}\cdot \vect{\hat{N}_{\mathcal{K}}}} \ket{\Psi_\mathrm{out}}\\
    ~&=\bra{\Psi_\mathrm{in}} \hat{U}^\dagger e^{i \vect{\eta}\cdot \vect{\hat{N}_{\mathcal{K}}}} \hat{U}\ket{\Psi_\mathrm{in}}
\end{align}
Written in this form, $x({\vect{\eta}})$ can be interpreted as the amplitude of $\ket{\Psi_\mathrm{in}}$ staying intact through a virtual interferometer $$\op{V}(\vect{\eta}) = \hat{U}^\dagger e^{i \vect{\eta}\cdot \vect{\hat{N}_{\mathcal{K}}}} \hat{U}$$
To ease notation, we will denote $V(\vect{\eta})$ as simply $V$. Even though in the main part of the paper we consider an input state with one photon per mode, here we do a slightly more general derivation encompassing states of more than one photon per input mode, with mode occupation numbers given by a vector $\vect{r}$. Using the standard trick of inserting $\op{V} \op{V}^{\dagger}$ in between each creation operator, we obtain
\begin{align}
    x({\vect{\eta}}) &= \frac{1}{\mu(\vect{r})} \braAket{0}{\prod _{j=1}^n  \op{a}_{d_j(\vect{r}), \phi_{d_j(\vect{r})}}  \prod _{j=1}^n \sum _{k=1}^m
    V_{d_j(\vect{r}),k} \op{a}^\dagger_{k, \phi_{d_j(\vect{r})} }}{0}\\
    & = \frac{1}{\mu(\vect{r})} \sum _{k_1, \dots, k_n=1}^m \left(\prod _{j=1}^n V_{d_i(\vect{r}),k_i} \right) \braAket{0}{\prod _{j=1}^n  \op{a}_{d_j(\vect{r}), \phi_{d_j(\vect{r})}}  \op{a}^\dagger_{k_j, \phi_{d_j(\vect{r})} }}{0}.\\
\end{align}
Let's now compute the quantity between brackets by expanding the internal degrees of freedom into a basis
\begin{equation}
\ket{\phi_j} = \sum _\alpha ^m c_{j\alpha} \ket{\Phi_\alpha}. 
\end{equation}
As there are only $n$ photons, we could stop the sum to $n$ as the other coefficients will be zero, but to ease the notation we keep the sum running up to $m$ with the understanding that some coefficients are zero by construction. Thus
\begin{align}
    \braAket{0}{\prod _{j=1}^n  \op{a}_{d_j(\vect{r}), \phi_{d_j(\vect{r})}}  \op{a}^\dagger_{k_j, \phi_{d_j(\vect{r})} }}{0} &=\sum _{\alpha_1, \dots, \alpha_n, \beta_1, \dots, \beta_n = 1}^m c_{d_1, \alpha_1}^* \dots c_{d_n, \alpha_n}^* c_{d_1, \beta_1} \dots c_{d_n, \beta_n} \\
    &\times \braAket{0}{\prod _{j=1}^n  \op{a}_{d_j(\vect{r}), \Phi_{\alpha_j}}  \op{a}^\dagger_{k_j, \Phi_{\beta_j} }}{0}
\end{align}
This last quantity is given by
\begin{equation}
\braAket{0}{\prod _{j=1}^n  \op{a}_{d_j(\vect{r}), \Phi_{\alpha_j}}  \op{a}^\dagger_{k_j, \Phi_{\beta_j} }}{0} = \sum_{\sigma \in S_n} \prod _{i=1}^n \delta_{d_i(\vect{r}), k_{\sigma_i}} \delta_{\alpha_i, \beta_{\sigma_i}}.
\end{equation}
Plugging into the expression for $x({\vect{\eta}})$ we obtain
\begin{align}
     x({\vect{\eta}}) &= \frac{1}{\mu(\vect{r})}  \sum _{\alpha_1, \dots, \alpha_n}^m \sum_{\sigma \in S_n}\prod _{j=1}^n V_{d_i(\vect{r}),d_{\sigma^{-1}_i}(\vect{r})} c_{d_i, \alpha_i}^* c_{d_i, \alpha_{\sigma^{-1}_i}}
\end{align}
where we use the fact that if $\sigma \in S_n$ then $\prod_{i = 1}^n A_{i, \sigma_i} = \prod_{i = 1}^n A_{\sigma^{-1}_i, i} $ to rearrange the product. Next, we eliminate the sums over $\alpha's$ by recombining the coefficients $c_{j, \alpha}$ into scalar products of the wavefunctions of internal degrees of freedom as
\begin{equation}
\braket{\phi_i}{\phi_j} = \sum_\alpha c_{i, \alpha}^* c_{j, \alpha} \equiv S_{ij}.
\end{equation}
This leads to the expression
\begin{align}
     x({\vect{\eta}}) &= \frac{1}{\mu(\vect{r})} \sum_{\sigma \in S_n}\prod _{j=1}^n V_{d_i(\vect{r}),d_{\sigma^{-1}_i}(\vect{r})} S_{d_i(\vect{r}),d_{\sigma^{-1}_i}(\vect{r})} 
\end{align}
By defining, as is conventional, the following reduced matrices to lighten the notation
\begin{align}
    \tilde{M}_{ij} &= V_{d_i(\vect{r}),d_{j}(\vect{r})}\\
    \tilde{S}_{ij} &= S_{d_i(\vect{r}),d_{j}(\vect{r})}
\end{align}
we recover a compact expression 
\begin{align}
     x({\vect{\eta}}) &= \frac{1}{\mu(\vect{r})} \sum_{\sigma \in S_n}\prod _{j=1}^n \tilde{S}_{i, \sigma_i}\tilde{M}_{i, \sigma_i} \\
     &= \frac{1}{\mu(\vect{r})} \perm{\tilde{S} \odot \tilde{M}}
\end{align}
involving a single $n\times n$ permanent of the elementwise (Hadamard) product between two matrices: the first constructed from entries of the Gram matrix of the internal degrees of freedom and the second from the interferometer $V$. When considering an input of photons occupying modes $(1,\dots,n)$, this reduces to the expression presented in the main text.

\section{Binned distributions of Fourier interferometers}
\subsection{Explicit probabilities for a single subset}
In order to derive the expressions for the single-mode distributions obtained in Sec.~\ref{sec:single_mode_Fourier}, we first derive a general expression for the photon-number distribution in a single subset of output modes which may be of independent interest. First we note that we can write Eq.~\eqref{eq:defV} as 
\begin{equation}
    V(\eta)= \left(\mathds{1}+(e^{i\eta}-1)H\right), 
\end{equation}
where the matrix $H$ is defined in terms of the interferometer $U$ as  
\begin{equation}
    \label{eq:H_Shchesnovich}
    H_{a,b} = \sum _{l \in \mathcal{K}} U_{l,a}^* U_{l,b},  
\end{equation}
for some subset of interest denoted as $\mathcal{K}$. Let us define $H_n$ as the $n\times n$ submatrix obtained from the first $n$ rows and columns of $H_n$ and $H_n'=\Smatrix\odot H_n$. Using the fact that the diagonal elements of the $\Smatrix$ matrix are $\Smatrix_{ii}=1$ we can write 
\begin{align}
\Smatrix\odot V_n(\eta)&= \Smatrix\odot \left(\mathds{1}_n+(e^{i\eta}-1) H_n\right)\\
&= \mathds{1}_n+(e^{i\eta}-1)H'_n
\end{align} 
Using an identity from Minc \cite{minc} (Chapter 2.2 exercise 5), the expression for the amplitudes $x({\nu_l})$, with $\nu_l=2\pi l/(n+1)$ can be expanded as follows
\begin{align}
x(\nu_l)&=\perm{\mathds{1}_n+(e^{i\nu_l}-1)H'_n}\\
&=1+\sum_{a=1}^{n}c_a (1-e^{i \nu_l})^a.\label{eq:expansion_perm}
\end{align}
The coefficients $c_a$ are given by 
\begin{equation}\label{eq:ca}
c_a=(-1)^a \sum_{\bar{w}\in Q_{a,n}}\perm{H'_n[\bar{w}]},
\end{equation}
where $Q_{a,n}$ denotes the set of all strictly ordered subsets of $\bar{w}\subset\{1,2,\dots,n\}$ containing $a$ elements. Furthermore, $H'_n[\bar{w}]$ denotes an $a\times a$ submatrix of $H'_n$ whose rows and columns are picked according to $\bar{w}$.  Plugging in Eq.~\eqref{eq:expansion_perm} into Eq.~\eqref{eq:Pk_from_xl} we can obtain, after some manipulations, an explicit expression for the probabilities of observing $k$ photons in the subset $\setparti$
\begin{equation}\label{eq:pk_expansion}
P(k)= (-1)^k\sum_{a=k}^n {{a}\choose{k}} c_a. 
\end{equation}
This expression is of limited use since in general it is given by a sum of exponentially many permanents. However, in some particular cases (such as the one in Sec.~\ref{sec:single_mode_Fourier}) it can be used to obtain analytical results for the probabilities. Moreover, it can be used to recover some results that were previously obtained. In particular, it can be seen from this expression that the probability that all photons end up in the chosen subset, which can be seen as a generalized bunching probability, is given by 
\begin{equation}
P(n)=\perm{H'_n}, 
\end{equation}
retrieving the result derived in \cite{shchesnovich2016universality}. It is also possible to see from the probability of not observing any photons in subset $\mathcal{K}$ is given by 
\begin{equation}
P(0)=\perm{\mathds{1}_n-H'_n}. 
\end{equation} 
The latter expression is consistent with the fact that this probability is the same as that of seeing $n$ photons in the complement of subset $\mathcal{K}$ and was considered in \cite{shchesnovich2021distinguishing}. Therein, it was also shown that this quantity can be used to distinguish certain efficient classical simulation algorithms from  boson samplers with partially distinguishable inputs for constant density $\rho=n/m$.

\subsection{Single mode output distribution}\label{secapp:single_mode_FT}
We now apply our previous derivation to the problem of obtaining the photon number distribution in a single detector of the Fourier interferometer. Without loss of generality we choose the subset $\setparti_1=\{1\}$. In this case, for an $m$-mode Fourier interferometer the $H$ matrix takes the simple form
\begin{equation}
H_{i,j}= F^*_{ 1 i}F_{1 j}= \frac{1}{m}. 
\end{equation}
For indistinguishable photons we have that $H'_n= H_n$ and the coefficients $c_a$ from Eq.~\eqref{eq:ca} are given by 
\begin{align*}
c_a&= (-1)^a \sum_{\bar{w}\in Q_{a,n}}\perm{H_n[\bar{w}]}\\
&= (-1)^a \sum_{\bar{w}\in Q_{a,n}} \frac{a!}{m^a}\\
&= (-1)^a {{n}\choose{a}} \frac{a!}{m^a}, 
\end{align*}
where we used the fact that the number of possible strictly order subsets from $Q_{a,n}$ is $|Q_{a,n}|={{n}\choose{a}}$. Plugging in Eq.~\eqref{eq:pk_expansion}, we obtain 
\begin{equation}
P_k^B= \sum_{a=k}^n (-1)^{k+a} {{a}\choose{k}} {{n}\choose{a}} \frac{a!}{m^a}. 
\end{equation}
Although to our knowledge this expression has not been derived before, it is known from the general results of Ref.~\cite{cushenhudson1971} that the single mode density matrix converges to a thermal state with an average number of photons $n/m$. In contrast, it can be seen that if we have fully distinguishable particles at the input, the distribution $P_k^D$ is a binomial distribution. In this case, the probability that each particle appears in the first output mode is $1/m$ and hence the probability of observing $k$ particles in this mode is 
\begin{equation}
P_k^D= {{n}\choose{k} }\frac{1}{m^k}\left(1-\frac{1}{m}\right)^{n-k}.
\end{equation}
In the regime of constant density $\rho=n/m= \text{const.}$, this probability distribution tends to a Poisson distribution with mean $\rho$. It can be seen that these two distributions are sufficiently far apart and that it should be possible to distinguish them efficiently, i.e.~with a number of samples growing polynomially in $n$, both in the constant density regime and in the collision free regime $m=n^2$.

\subsection{Photon number distribution in the odd modes}\label{secapp:special_subset}
We will now restrain ourselves to the case where the number of modes $m$ is equal to the number of input photons $n$ and thus consider exactly one photon per input mode. We will show that for the choice of a specific partition of the output modes, a striking difference exists between the distinguishable and indistinguishable cases by computing analytically those probability density functions. Namely, we will consider the special subset of the output modes consisting of the modes with odd numbers $\setparti = \{1,3,5,\dots,n-1\}$ and will restrict ourselves to $n$ even. In this scenario, we show that if the input are bosons the probability of seeing an odd number of photons in $\setparti$ is $0$, whereas the probability of seeing an even number of photons follows a simple binomial distribution. We start by computing 
\begin{align}
     x({\vect{\eta}}) &= \perm{F^\dagger \Lambda(\eta) F}. 
\end{align}
We will first simplify the matrix product inside the permanent. We start by writing 
$$ \Lambda(\eta) = \mathds{1} + \sum _{j\in \setparti} \left(e^{i\eta} -1\right) \ket{j}\bra{j}$$
so that 
\begin{align*}
     &(F^\dagger \Lambda(\eta) F)_{ab} =
      \delta _{ab} + \left(e^{i\eta} -1\right) \sum _{j\in \setparti} F^\dagger_{aj} F_{jb}\\
    &= \delta _{ab} + \left(e^{i\eta} -1\right) \sum _{j\in \setparti} \frac{1}{n} \exp \left(\frac{2\pi i}{n} (j-1)(a-b) \right) \\
    &= \delta _{ab} + \left(e^{i\eta} -1\right) \frac{1}{n} \sum _{k = 0}^{n/2 -1} \exp \left(\frac{4\pi i}{n} k(a-b) \right) \\
   \end{align*} 

Now, let's note that the sum is a geometric series 
\begin{align*}
    &\sum _{k = 0}^{n/2 -1} \exp \left(\frac{4\pi i}{n} k(a-b) \right) =\\ &=\begin{cases} \frac{n}{2}, & \mbox{if } |a-b| = 0 \mbox{ or } \frac{n}{2} \\
    0 & \mbox{otherwise}
    \end{cases}
\end{align*}
given that $a,b$ are integers.
Thus 
\begin{align}
(F^\dagger \Lambda(\eta) F)_{ab} = \begin{cases} \frac{e^{i\eta} +1}{2} & \mbox{if } a = b \\
\frac{e^{i\eta} -1}{2} & \mbox{if } |a-b| = \frac{n}{2} \\
0 & \mbox{otherwise}
\end{cases}
\end{align}
In other words, 
\begin{align}
&V(\eta)=F^\dagger \Lambda(\eta) F = \\
&=\mbox{circ}\left(\frac{e^{i\eta} +1}{2}, 0,\dots,0, \frac{e^{i\eta} -1}{2}, 0,\dots,0   \right)
\end{align}
where $\text{circ}(\vec{v})$ denotes the circulant matrix whose first row is $\vec{v}$.
Now let's see that this form allows us to compute the permanent analytically. We have that
\begin{equation}
\perm{V(\eta)}= \sum_{\sigma\in S_n} \prod_i V(\eta)_{i \sigma(i)}.
\end{equation}
Due to the large number of zeros in $V(\phi)$ the sum over all permutations can be greatly simplified. In fact, the only permutations that lead to a non-zero product $\prod_i V(\eta)_{i \sigma(i)}$ are those such that $\sigma(i) =i$ or $\sigma(i) = i+n/2~~ (\text{mod}~n)$. Such permutations can be written as a product of disjoint permutations
\begin{equation}
\sigma= \gamma_1\gamma_2\dots\gamma_{n/2}
\end{equation} 
where $\gamma_i\in \{e, \tau_i \}$, $e$ is the identity permutation and 
$\tau_i$ is the transposition that switches elements $i$ with element $i+n/2~~ (\text{mod}~n)$. Hence we can write 
\begin{align}
\perm{V(\phi)}&=
\sum_{\gamma_1}\sum_{\gamma_2}\dots\sum_{\gamma_{n/2}}
\prod_{i=1}^{n/2} V(\eta)_{i, \gamma_i(i)} V(\eta)_{\frac{n}{2}+i, \gamma_i(\frac{n}{2}+i)}\\
&=\prod_{i=1}^{n/2}\sum_{\gamma_i} V(\eta)_{i, \gamma_i(i)}V(\eta)_{\frac{n}{2}+i, \gamma_i(\frac{n}{2}+i)}\\
&=\prod_{i=1}^{n/2}\left(\left(\frac{e^{i\eta} +1}{2}\right)^2+\left(\frac{e^{i\eta} -1}{2}\right)^2 \right)\\
&=\left(\left(\frac{e^{i\eta} +1}{2}\right)^2+\left(\frac{e^{i\eta} -1}{2}\right)^2 \right)^{n/2}\\
&= \frac{1}{2^{n/2}} (e^{2i\eta} +1)^{n/2}
\end{align}
In the second step we used the fact that the $\gamma_i$'s are disjoint permutations to switch the product with the sum. Finally, by applying Eq.~\eqref{eq:Pk_from_xl} we obtain 
\begin{equation}
P_k^B = \begin{cases}
0 &\mbox{if }k \mbox{ is odd}\\
\frac{1}{2^{n/2}}{n/2 \choose k}&\mbox{if }k \mbox{ is even}
\end{cases}
\end{equation}

Comparatively, we can find the counterpart of this expression for distinguishable particles following a simple statistical argument. For an unbiased interferometer (such as the Fourier interferometer), the probability that a given particle falls into the partition A is $q = |A|/m$ such that the probability to find $j$ photons inside the partition is given by the binomial distribution 
$$
P_j^{D} = {n \choose j}q^j (1-q)^{n-j}
$$
which, in the case discussed above, reduces to 
$$
p_j^{D} = \frac{1}{2^n}{n \choose j}.
$$
The striking difference between the two cases is represented in Fig.~\ref{fig:FourierAodd}. It is possible to see that the TVD between these two probability distributions tends to a constant as $n$
goes to infinity.

\section{Further numerical investigations}

In this appendix, we present extra numerical evidence to support the claims made in the main text. All plots of this paper are reproducible  by using the associated project, \textsc{BosonSampling.jl} which contains the exact code needed for their execution. The package also allows to explore different parameter ranges.

\label{appendix:numerics}

\subsection{Default partition choice}

\label{appendix:equipartition}

Unless otherwise specified, the choice of partition is made as follows. We divide all physical output modes $m$ in $\sizeparti$ bins. For a lossy boson sampler, all the environment modes are always grouped together in a single bin. If $\sizeparti$ divides $m$, each physical bin is contains $m/\sizeparti$ modes. Otherwise, we can decompose $m = p (\sizeparti - 1) + q$. Then the first $\sizeparti - 1$ bins are of size $p$ while the remaining one of size $q$. We always take consecutive modes in building each bin. This choice is not very important as we approximately obtain the Haar average value of the probabilities by computing them for several Haar random unitaries. We remark that for the Haar averaged probabilities only the size of each bin matters and not the specific choice of partitionas. Arbitrary arrangements for the bins, which may be useful to test particular interferometers, can readily be simulated with the provided codes.

\subsection{Variability of the TVD}

\begin{figure}[h!]
  \centering
  \includegraphics[width=0.5\textwidth]{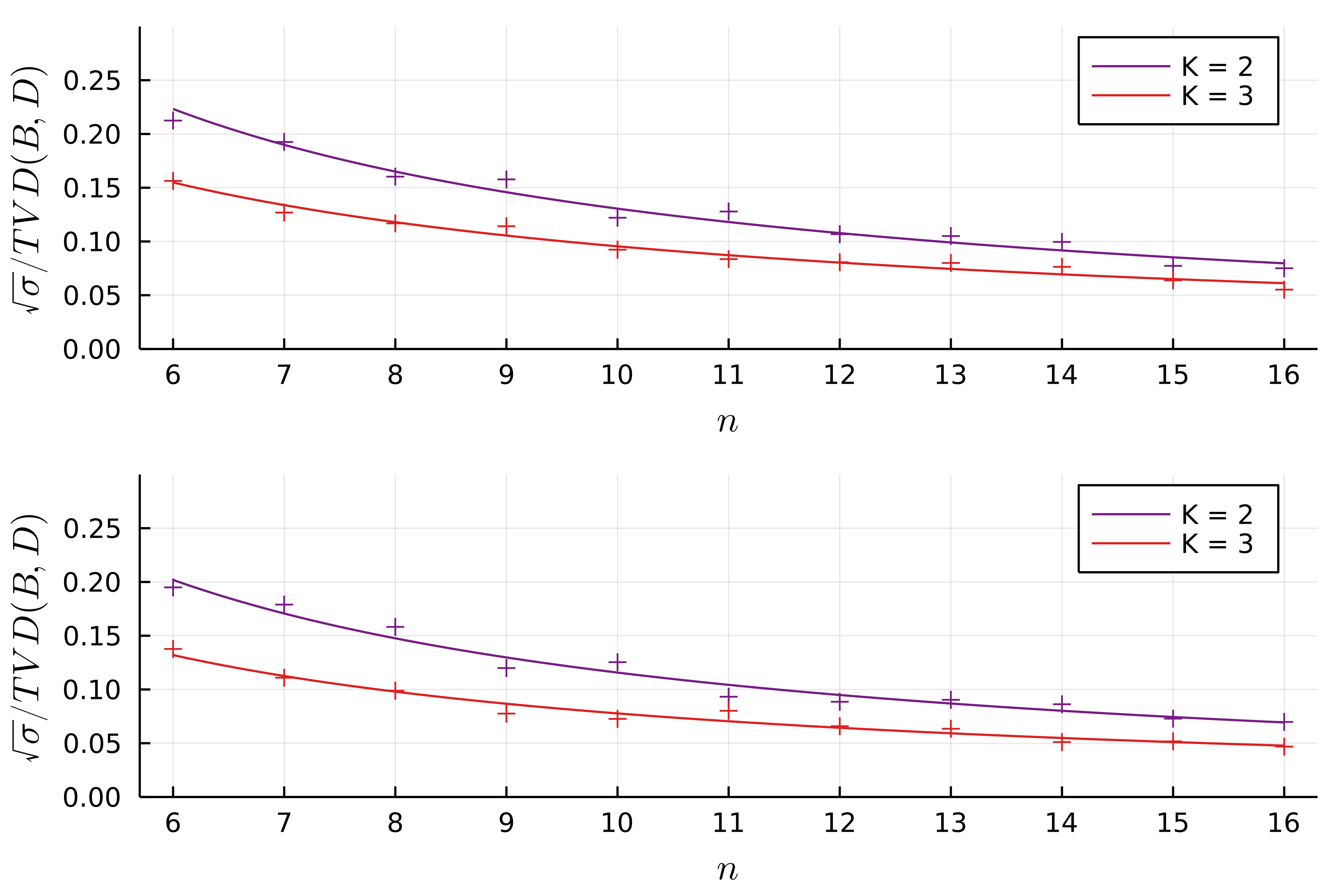}
  \caption{\textbf{Variability of the TVD with system size.}  We plot the same parameters as in Fig. \ref{fig:size} but instead of looking at the TVD we display its coefficient of variation (CV), defined as the standard deviation divided by the mean.
  To a good approximation, it follows a power law of form $CV \propto n^{-1}$ in all plots with a slightly varying prefactor of order one.
    }
  \label{fig:coefficient_variation_size}
\end{figure}

In Fig. \ref{fig:coefficient_variation_size}, we show the behaviour of the variance bars of Fig. \ref{fig:size}, showing how the TVD converges to its mean value. 

\subsection{TVD with partial distinguishability}

\begin{figure}[h!]
  \centering
     \includegraphics[width=0.5\textwidth]{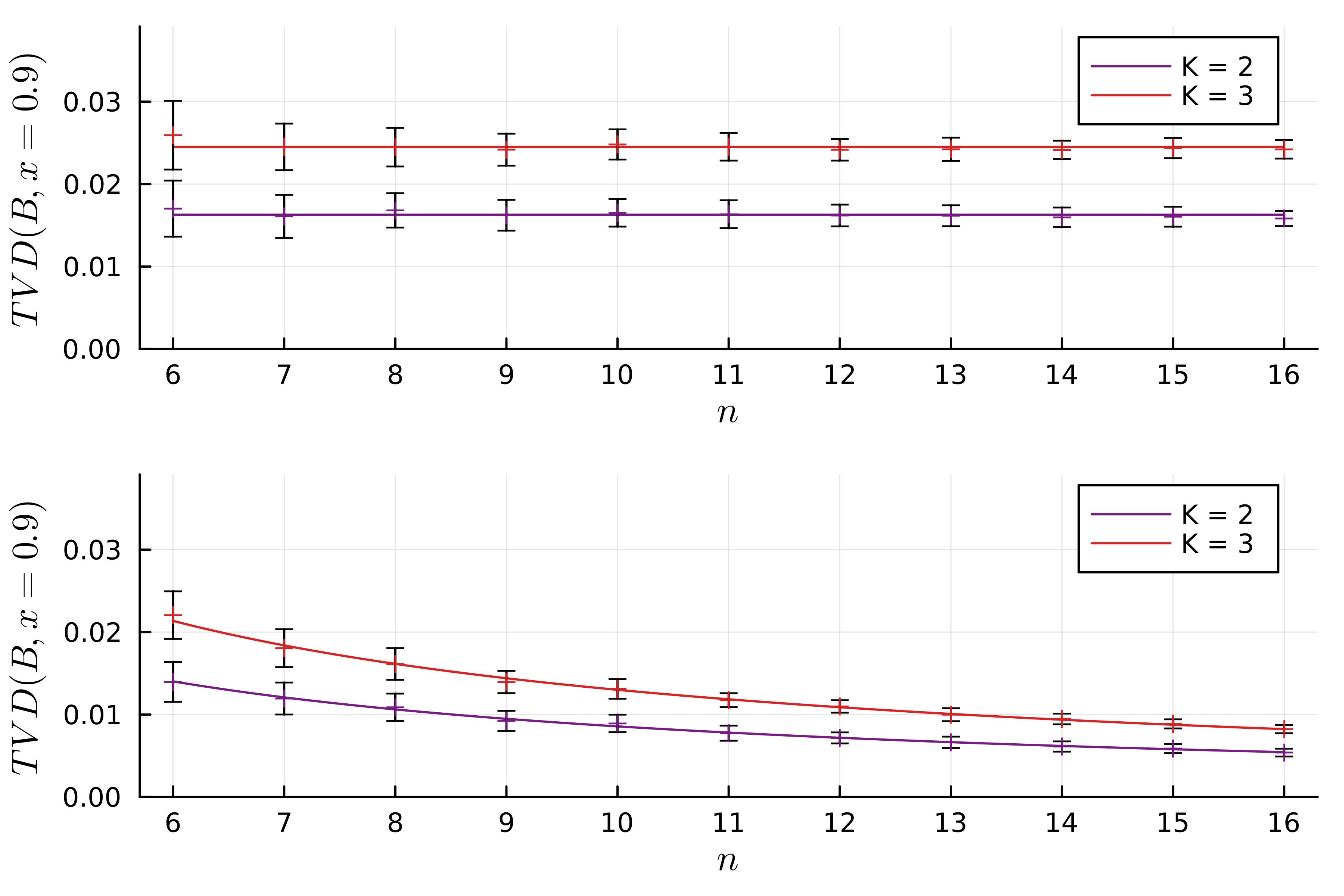}
  \caption{\textbf{TVD between bosonic and nearly indistinguishable particles.} This figure shows that the results shown in Fig. \ref{fig:size} hold even when comparing indistinguishable photons to nearly indistinguishable ones. All parameters are identical to those from Fig.~\ref{fig:size}, except from the fact that we compare ideal bosons to partially distinguishable ones with $x = 0.9$ (instead of $x = 0$). We confirmed this trend in multiple numerical trials with other values of distinguishability parameter $x$.} 
    
  \label{fig:sizeWithX}
\end{figure}

In Fig.~\ref{fig:sizeWithX}, we show here that the TVD decreases in a similar fashion than described in Fig. \ref{fig:size} when comparing indistinguishable particles to nearly indistinguishable ones. This is an important observation as this gives strong evidence for the scalability of our validation protocol even in real-world use cases. In addition, the fact that the distance between distributions may allow for the estimation of the distinguishability levels in the experiment, by finding the distinguishability parameter that gives the highest agreement with the experimental data.

\subsection{The role of boson density}

\begin{figure}[h!]
  \centering
  \includegraphics[width=0.5\textwidth]{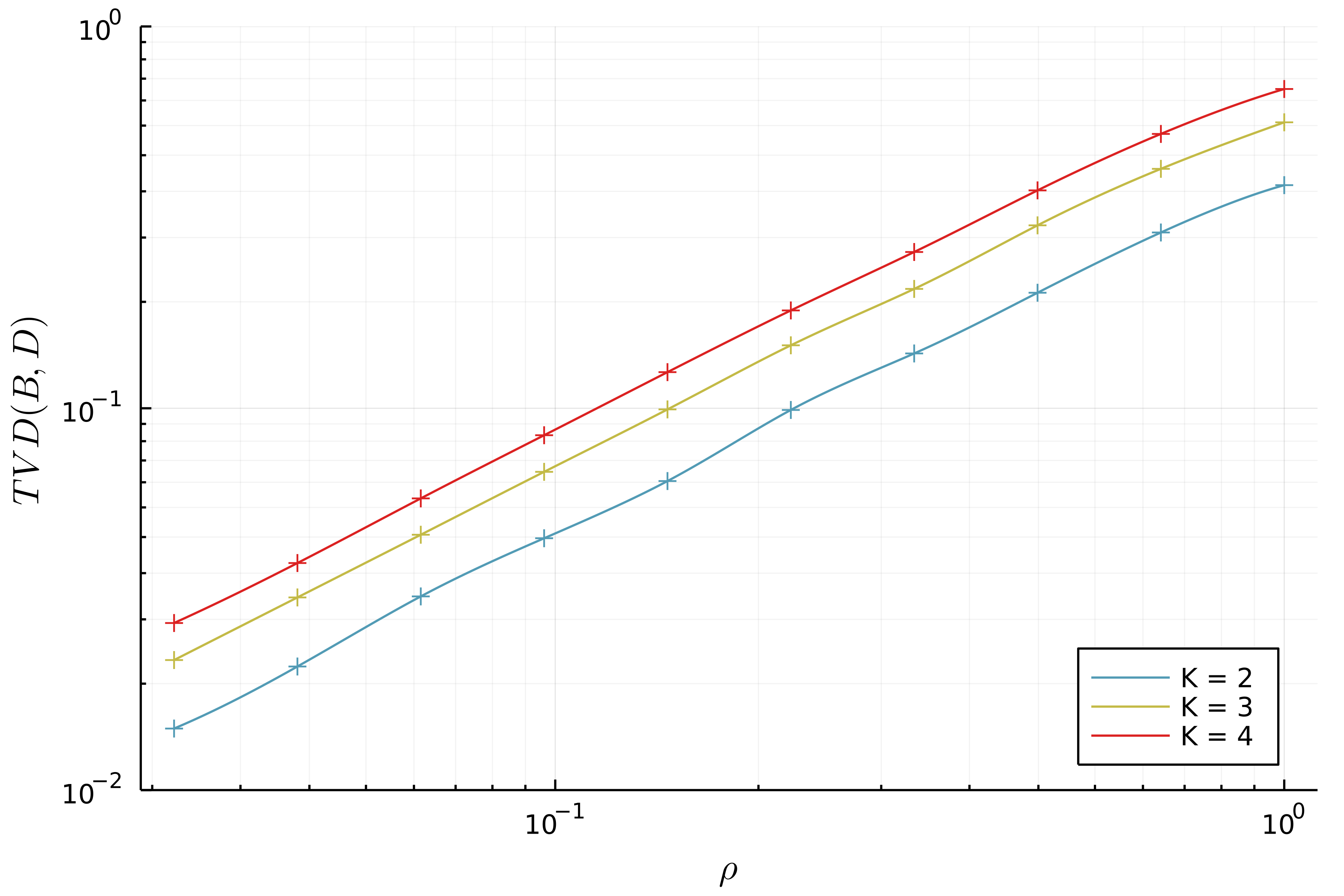}
  \caption{\textbf{Effect of boson density.} The Haar-averaged TVD between bosonic and distinguishable particles (asymptotically) depends only on the boson density $\rho$, see \eqref{eq:haarAveragePartitions}. We show how it evolves in the example of $n=10$ and $m=10,\dots,300$ with a number of subsets $K=2,3,4$. Real-world experiments typically have a high boson density, for instance \cite{wang2019boson} has $n=20$ and $m=60$ thus $\rho = 1/3$. Note that the hardness of boson sampling is shown in a very dilute case of $m \propto n^6$ and conjectured in the "no collision" regime $m \propto n^2$ \cite{bosonsampling}.}
  \label{fig:density}
\end{figure}

\begin{figure}[h!]
  \centering
  \includegraphics[width=0.5\textwidth]{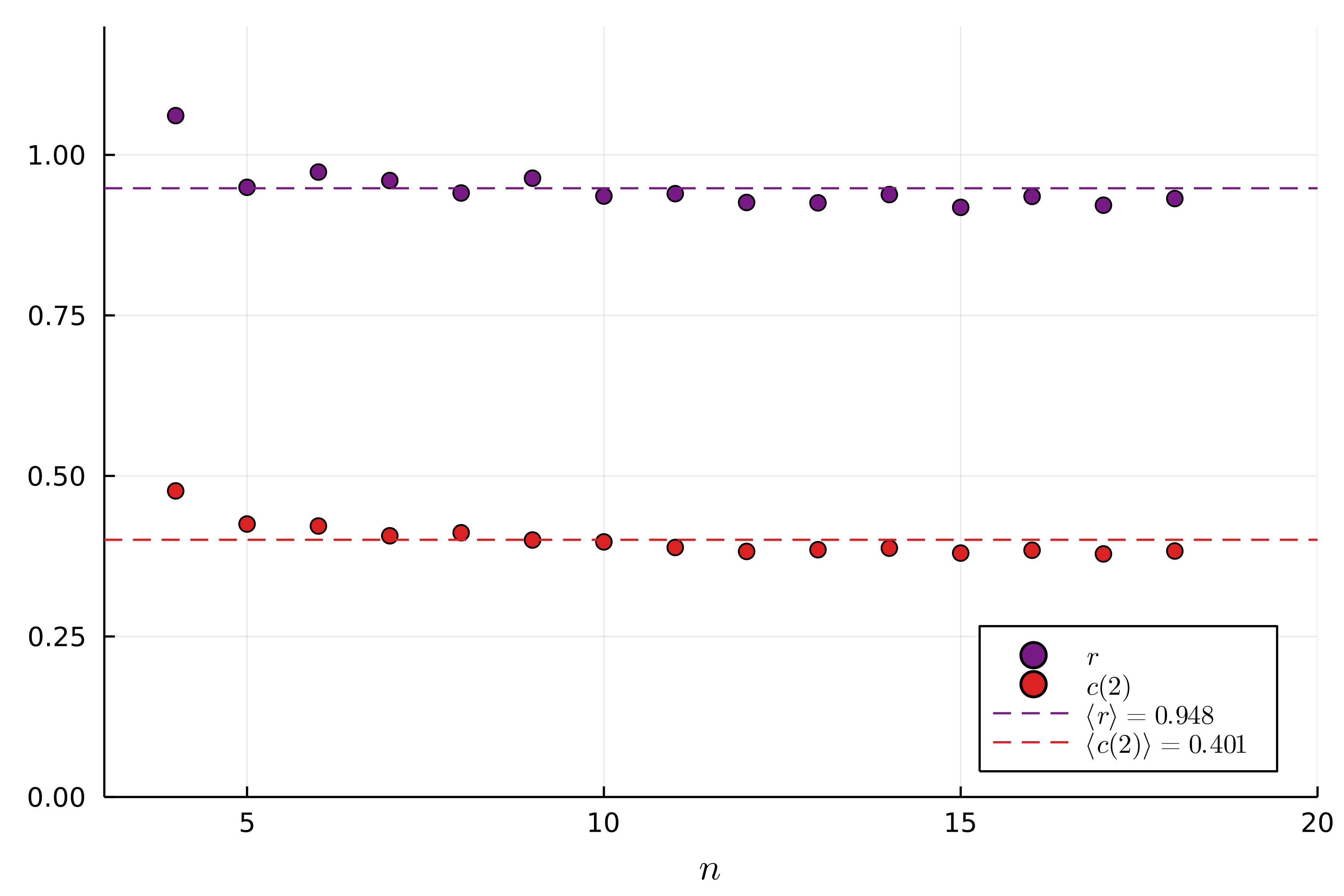}
  \caption{\textbf{Validity of the power law.} In Fig \ref{fig:density} we claim that the density is the prime factor modifying the TVD.
  Here we provide a justification for this by looking at the power-law fit $\tvd{B,D} = c(K,0)\rho^r$ with $K$ the number of subsets in the partition.
  We plot the coefficients for various values of $n$.
  The values of $m$ are such that the density ranges from $1,\dots,0.03$. $10$ values of $m$ are taken and are equally distributed in logarithm (such as seen on Fig. \ref{fig:density}).
  For each point, $100$ iterations are performed.}
  \label{fig:power_law_validity}
\end{figure}

\begin{figure}[h!]
  \centering
  \includegraphics[width=0.5\textwidth]{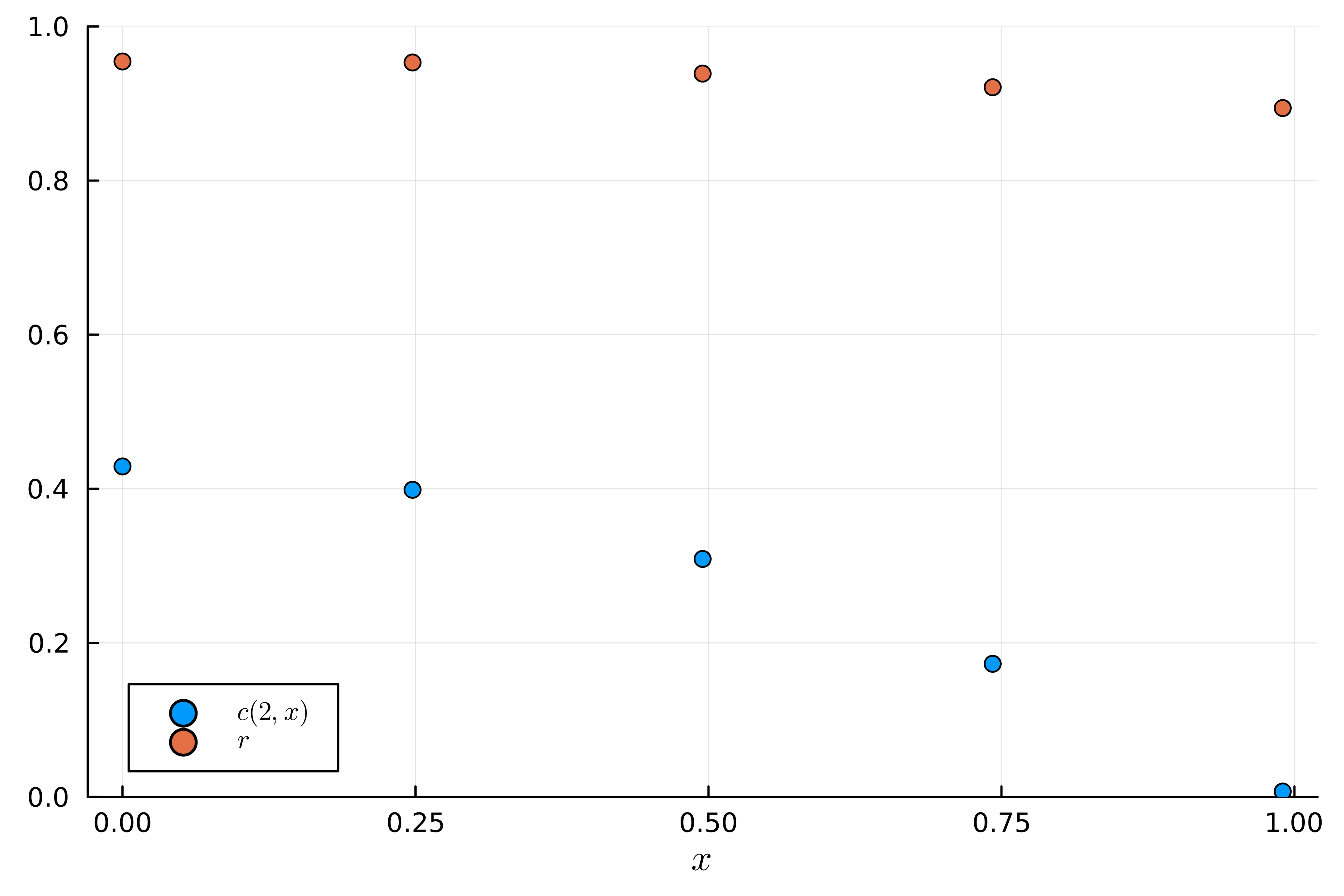}
  \caption{\textbf{Validity of the power law for various values of $x$.}. We show the evolution of the coefficient $c(2,x)$ as well as the power $r$ for $n = 8$ over various values of $x$. We see small variations in the $r$ but the overall behaviour seems to hold well.
  For each point, $100$ iterations are performed. The last point on the right has value $x=0.99$.}
  \label{fig:power_law_validity_x}
\end{figure}

Our numerical investigations highlight the preponderant role of boson density $\rho = n/m$ in the behavior of the TVD between binned distributions of different types of input particles. This dependency is shown in Fig. \ref{fig:density} for the TVD between distinguishable and indistinguishable photons. These power law fits at low number of photons are useful to understand the scalability of the validation technique we consider for larger experiments.
Overall, we find that 
\begin{equation}
    \label{eq:powerLawDensityAppendix}
    \tvd{B,x} \approx c(K,x) \rho
\end{equation}
with $c(2, 0) \approx 0.41$, $c(3, 0) \approx 0.67$. We see how the above equation holds when using different values of $n$ in Fig. \ref{fig:power_law_validity}. While some slight variation is found in small sized systems, the power law seems to become nearly independent of $n$ in larger systems. Likewise, we also check that it holds to a good approximation when comparing indistinguishable photons to some with partial distinguishability $x$ in Fig. \ref{fig:power_law_validity_x}.

\subsection{Dependency of number of samples on density}

In this subsection, we give some details about Eq.~\ref{eq:number_samples}. We use as ansatz the  following equation describing the dependence of the number of samples to validate the ideal boson sampler hypothesis against a model with partial distinguishable photons:
\begin{equation}
    n_s \approx \frac{d(K,x)}{\rho^{r}}
\end{equation}
The value of $d(K,x)$ depends on the partition size and the value of partial distinguishability. For $x=0$ we obtain $d(2,0) = 17.9$ and the power fit gives an exponent of $r = 2.53$ when using $n=10$ input photons. Our numerical evidence suggests that the fit becomes better and better as the density decreases, as finite size/density effects may be at play. The same can be said for the number of input photons. We also obtained preliminary evidence suggesting that the power law holds for other values of partial distinguishability, although a complete analysis is out of the scope of the paper. More precise numbers can be obtained using the publicly available codes (see Code Availability).

\subsection{Spoofing}\label{appendix:spoofing}

An open question left by our work is whether an adversary can spoof the validation test based on binned distributions and successfully pretend to generate samples from an ideal boson sampler, for example, by constructing a clever mock-up sampler based on an efficient classical algorithm. In this subsection, we study whether an adversary can easily reproduce the binned distributions obtained from an ideal boson sampler, without the knowledge of the choice of partitions used to validate the sampler. To do so, we quantify numerically in two different ways how much the binned distribution changes over different random choices of the bins, showing there is a significant variability in the distributions between different choices. 

\begin{figure}[h!]
  \centering
  \includegraphics[width=0.7\textwidth]{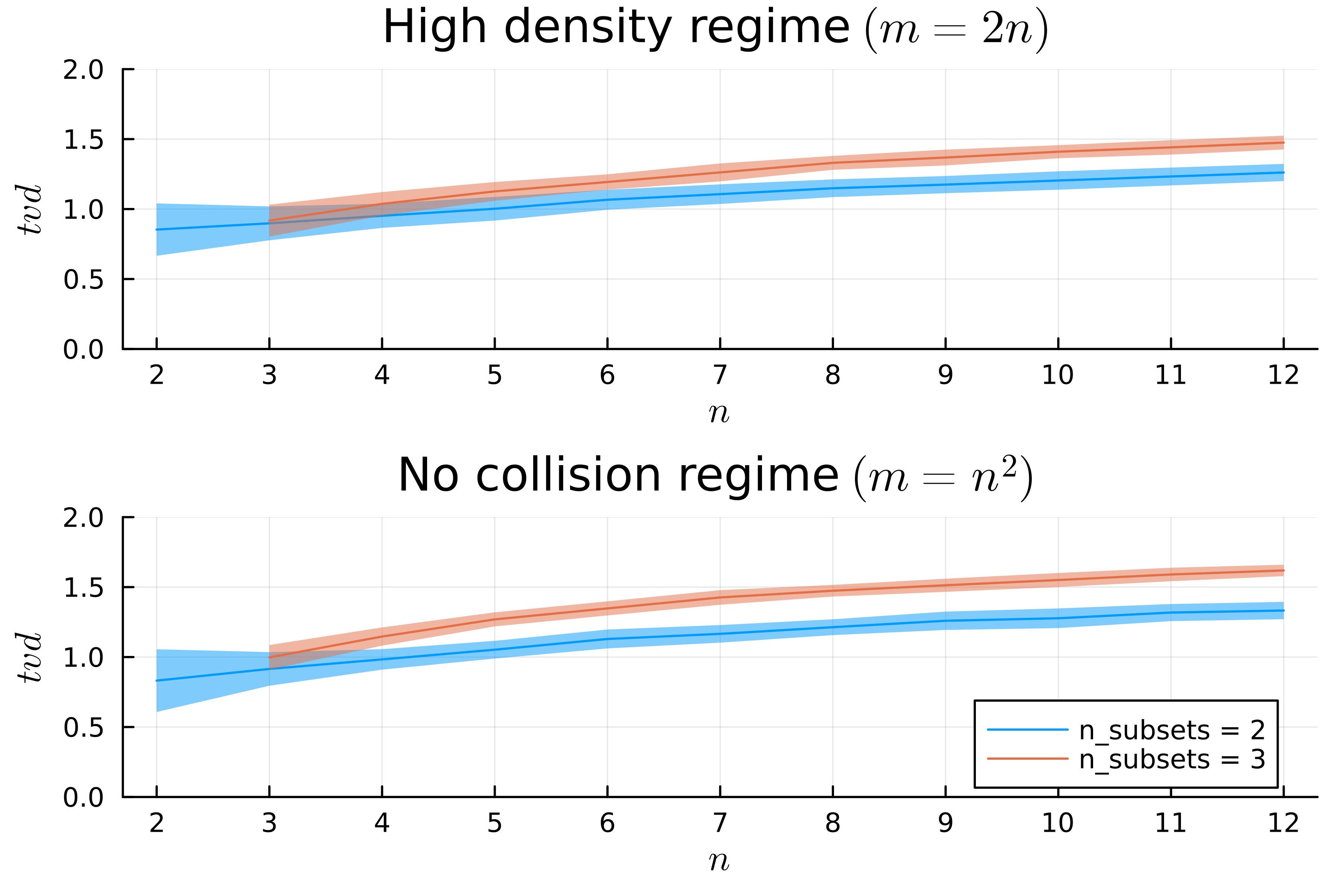}
  \caption{\textbf{Distance between randomly selected partitions with subsets of possibly different sizes. }  The same procedure is repeated $100$ random bin choices for each unitary. We then average, for each value of $n$, over $100$ Haar random unitaries. Two scenarios are presented: a high density regime, where $m=2n$ and a no collision regime with $m = n^2$. For both cases, we study the effect of choosing $2$ or $3$ bins.  }
  \label{fig:spoofing}
\end{figure}

First we consider a unitary drawn at random according to the Haar measure and compute the total variation distance between two random partitions with the same number of bins, but not necessarily the same sizes for each bin. An example for $m = 4$ could be to consider subsets of modes $\{1,2\}, \{3,4\}$  for the first partition while taking $\{1\}, \{2,3,4\}$ for the second. In Fig. \ref{fig:spoofing} we plot the total variation distance between two randomly picked distributions, averaging over random partition choices and Haar random unitaries. The large values of the TVD observed are explained by the fact that there are large differences of the distributions with different bin sizes, as can be seen from the asymptotic expression in Eq.~\eqref{eq:asymptoticsB}. For example, in a bin of size $L$ we would expect to see on average $L n/m$ photons. 

\begin{figure}[h!]
  \centering
  \includegraphics[width=0.7\textwidth]{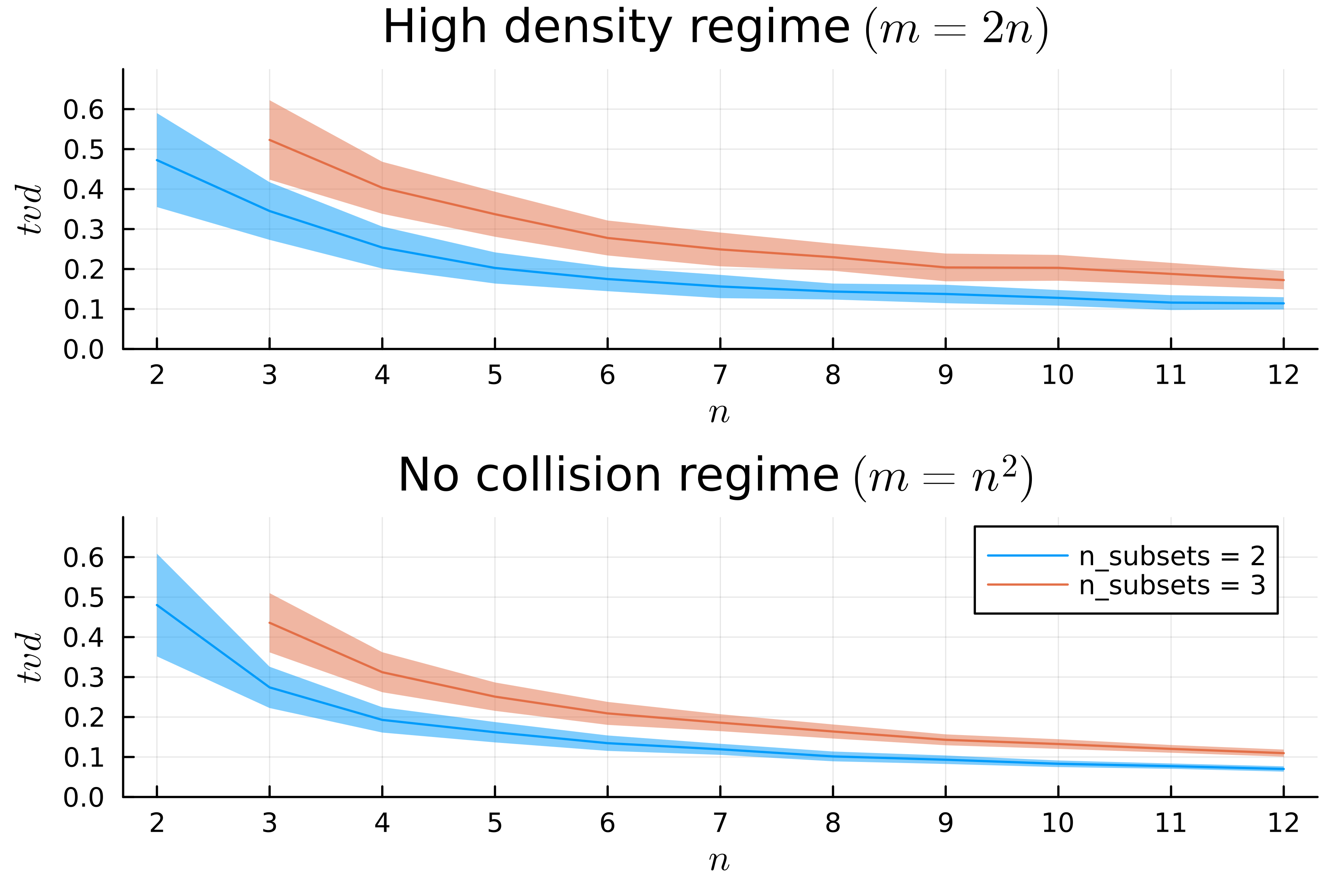}
  \caption{\textbf{Total variation distance between randomly selected partitions containing subsets of equal sizes.  }The same procedure is repeated $100$ random bin choices for each unitary. We then average, for each value of $n$, over $100$ Haar random unitaries. Two scenarios are presented: a high density regime, where $m=2n$ and a no collision regime with $m = n^2$. For both cases, we study the effect of choosing $2$ or $3$ bins. }
  \label{fig:spoofing_fixed_size}
\end{figure}

In our second example, we want to illustrate that even an adversary who is able to produce samples that correctly reproduce the Haar averaged properties of the binned distributions (e.g. by exploiting knowledge of Eq.~\eqref{eq:asymptoticsB}) would not be able to spoof the validation test.  To do so, in Fig. \ref{fig:spoofing_fixed_size}, we compare only binned distributions from randomly chosen subsets of fixed sizes. More precisely, in this second case, we divide the $m$ modes with $K$ equally sized subsets (up to the remainder of $m/K$). In this case, for $m=4$ we could for instance choose $\{1,2\}, \{3,4\}$ and $\{3,2\}, \{1,4\}$ (but not $\{1\}, \{2,3,4\}$ as the first subset does not have the same size as  $\{1,2\}$). 
In each case, we compute the TVD between the binned distributions resulting from these random bin choices, averaging over different random bin choices of equal sizes and also over Haar random unitaries. 

Even in this latter scenario, we see that the TVD between distribution decreases slowly with the number of photons. This suggests that with a polynomial number of samples, one could identify differences between a mock-up sampler that is able to reproduce the right Haar averaged properties of binned distributions and an ideal boson sampler. It would be interesting to analytically prove a polynomial decrease of the TVD in this scenario, a question we leave for future work.

\end{document}